\documentclass[usenatbib]{mnras}
\usepackage{amsmath}
\usepackage{newtxtext,newtxmath}
\usepackage[abs]{overpic}
\usepackage{xcolor}
\usepackage{float}                     % <<<<<
\usepackage{placeins}               % <<<<<
\usepackage[T1]{fontenc}
\DeclareRobustCommand{\VAN}[3]{#2}
\let\VANthebibliography\thebibliography
\def\thebibliography{\DeclareRobustCommand{\VAN}[3]{##3}\VANthebibliography}
\usepackage{graphicx}	% Including figure files
\usepackage{amsmath}	% Advanced maths commands
\newcommand{\BR}{Bright Rim}
\newcommand{\CCC}{\textit{CCC}}
\newcommand{\Lv}{Low--velocity}
\newcommand{\lv}{low-velocity}
\newcommand{\HFS}{Hub-Filament System}

\newcommand{\Hv}{\textit{High--velocity}}
\newcommand{\hv}{high-velocity}
\newcommand{\IF}{Ionisation Front}
\newcommand{\RRLs}{Radio Recombination Lines}
\newcommand{\SWS}{Spiders-Web System}

\newcommand{\SCL}{Shock-Compressed Layer}

\newcommand{\cc}{\,{\rm cm^{-2}}}
\newcommand{\ccc}{\,{\rm cm^{-3}}}
\newcommand{\duC}{\Delta u_{\mbox{\tiny CRIT}}}

\newcommand{\eightmum}{$8\,\upmu{\rm m}$}
\newcommand{\etan}{\eta_{\mbox{\tiny\it n}}}
\newcommand{\fPV}{f_{\mbox{\tiny PV}}}
\newcommand{\fion}{f_{\mbox{\tiny ION}}}
\newcommand{\gcc}{\,{\rm g\,cm^{-2}}}
\newcommand{\gccc}{\,{\rm g\,cm^{-3}}}
\newcommand{\HIIR}{H{\sc ii} Region}

\newcommand{\IO}{I_{\mbox{\tiny o}}}

\newcommand{\K}{\,{\rm K}}
\newcommand{\kms}{\,{\rm km\,s^{-1}}}
\newcommand{\kyr}{\,{\rm kyr}}
\newcommand{\MO}{M_{\rm o}}
\newcommand{\Mbar}{\mathcal{\bar{M}}}
\newcommand{\Msun}{\,{\rm M}_{_\odot}}
\newcommand{\Myr}{\,{\rm Myr}}
\newcommand{\mun}{\mu_{\mbox{\tiny\it n}}}
\newcommand{\muv}{\mu_v}
\newcommand{\NHM}{N_{_{\rm H_2}}}
\newcommand{\Ntot}{{\cal N}_{\mbox{\tiny tot}}}
\newcommand{\nCO}{n_{_{\rm CO}}}
\newcommand{\nCOp}{n_{_{\rm CO^+}}}
\newcommand{\nHM}{n_{_{\rm H_2}}}
\newcommand{\nHO}{n_{_{\rm H^o}}}
\newcommand{\nHp}{n_{_{\rm H^+}}}
\newcommand{\nel}{n_{_{\mbox{e}}}}
\newcommand{\npr}{n_{_{\mbox{p}}}}
\newcommand{\PDR}{Photon-Dominated Region} % {Photo-Dissociation Region}
\newcommand{\pc}{\,{\rm pc}}
\newcommand{\phin}{\phi_n}
\newcommand{\RO}{R_{\rm o}}
\newcommand{\RW}{R_{\mbox{\tiny W}}}
\newcommand{\RWn}{R_{\mbox{\tiny\it n}}}
\newcommand{\Rsink}{R_{\rm sink}}
\newcommand{\rhoO}{\rho_{\rm o}}
\newcommand{\rhosink}{\rho_{\rm sink}}
\newcommand{\seq}{\!=\!}
\newcommand{\sign}{\sigma_{\mbox{\tiny\it n}}}
\newcommand{\sigv}{\sigma_v}
\newcommand{\TO}{T_{\rm o}}
\newcommand{\tOB}{t_{\mbox{\tiny OB}}}

\newcommand{\thCRIT}{\theta_{\mbox{\tiny CRIT}}}
\newcommand{\twofourmum}{$24\,\upmu{\rm m}$}
\newcommand{\uO}{u_{\rm o}}

\title[]{Bipolar H\,\textsc{ii} regions produced by Cloud-Cloud Collisions}

\author[T. Georgatos et. al.]{
Theotokis Georgatos,$^{1,2}$\thanks{E-mail: publications@ras.ac.uk (KTS)}
Anthony P. Whitworth,$^{1,2}$ Richard W\"unsch$^3$ and Annie Zavagno$^{4,5}$
\\
$^{1}$Department of Physics and Astronomy, Cardiff University, Cardiff CF10 3AT\\
$^{2}$Royal Astronomical Society, Burlington House, Piccadilly, London W1J 0BQ, UK\\
$^{3}$ Astronomical Institute of the Czech Academy of Sciences,
Boční II 1401/1, 141 00 Praha 4, Czech Republic\\
$^{4}$ Aix Marseille Univ, CNRS, CNES, LAM, Marseille, France\\
$^{5}$ Institute Universitaire de France, 1 rue Descartes, Paris, France\\
}

\date{Accepted XXX. Received YYY; in original form ZZZ}

\pubyear{XXXX}

\begin{document}
\label{firstpage}
\pagerange{\pageref{firstpage}--\pageref{lastpage}}
\maketitle
%%%%%

% %%%%    Abstract of the paper
\begin{abstract}
We use numerical experiments to explore two possibilities: (i) that Bipolar \HIIR s are the result of Cloud-Cloud Collisions (CCCs), and (ii) that -- when allowance is made for the chaotic nature of such collisions, the short duration of the bipolar phase, and different viewing angles -- a large proportion of all \HIIR s might be the aftermath of CCCs. To reduce the parameter space, our experiments only consider head-on collisions between two $500\Msun$ clouds, with three different levels of turbulence, and two different collision velocities; the collision velocities define the `collision axis'. In all experiments OB stars only condense out after a \SCL\ has formed (perpendicular to the collision axis), and fragmented to produce a \HFS, with the OB stars forming in the Hub. Ionising radiation from the OB stars excites an \HIIR, which tends to expand more rapidly in directions close to the collision axis, and more slowly in directions orthogonal to the collision axis, where it encounters the dense gas of the \SCL. Consequently the \HIIR\ may appear bipolar, for a short period during its evolution, if observed at sufficiently large angle to the collision axis. Viewed from smaller angles, the waist appears as a \BR, similar to conventional approximately spherical \HIIR s. Under this circumstance there are other metrics -- based on the extent of  diffuse freefree emission, the velocity dispersion of \RRLs, dust emission at mid-infrared wavelengths -- that might indicate the aftermath of a CCC, and establish CCCs as a dominant trigger for high-mass star formation. % [246 words]
\end{abstract}
%%%%%

%%%%%
\begin{keywords}
stars: formation -- stars: massive -- ISM: clouds --ISM: HII regions-- ISM: kinematics and dynamics
\end{keywords}
%%%%%

%%%%%
\section{Introduction}
%%%%%

\textit{Herschel} observations have established that star formation is closely linked to filamentary structures \citep{molinari2010clouds, andre2010filamentary, Arzoumanian2011, konyves2015census, wang2015large, marsh2016census, howard2019, howard2021, hacar2023, wang2024filamentary}. In regions forming massive stars, these filaments are frequently organised into \HFS s \citep{deharveng2015bipolar, myers2009, kumar2020, samal2018bipolar, 2013Peretto, 2022Peretto}, and these highly anisotropic density fields must strongly influence the evolution of the \HIIR s excited by the massive stars. The ionising radiation may suppress star formation by dispersing dense gas \citep{Dale2017, Khullar2024}, or promote fragmentation by compressing neutral material at the boundary of the \HIIR\ \citep{Dale2012, Dale2013, elmegreen1977sequential, whitworth1994collect}. Understanding how ionising radiation propagates through structured molecular clouds is therefore essential for evaluating the role of massive stars in regulating cloud evolution and star formation.

Early theoretical work describing the evolution of \HIIR s assumed spherical symmetry \citep[e.g.][]{stromgren1939physical, Spitzer1978}. However, observed and modelled \HIIR s often deviate strongly from spherical symmetry, exhibiting cometary \citep{1985reid,2014Immer, steggles2017hydrodynamical}, blister \citep{whitworth1979, yorke1983, henney2005photoevaporation}, champagne-flow \citep{tenorio1979dynamics}, shell-like \citep{2020Dewangan}, and bipolar morphologies \citep{deharveng2015bipolar, comeron2018ionizing, samal2018bipolar, 2022larose}. Even modest density gradients can lead to strongly anisotropic ionisation fronts and highly directional outflows of ionised gas \citep{Peters2010, Geen2015, Mackey2015}. While spherical and blister-type \HIIR s have been studied extensively, Bipolar \HIIR s remain comparatively poorly explored, particularly with regard to their formation pathways, dynamical evolution, and observational appearance.

Bipolar \HIIR s are characterised by two lobes of ionised gas, joined at a constricted waist, and are commonly interpreted as the result of ionising radiation escaping preferentially along two opposing directions where the density is lower; by implication, expansion of the \HIIR\ is being inhibited by denser material in the other directions. Observational studies have identified many candidate bipolar \HIIR s, for example Sh 201 \citep{Deharveng2012} and G319.88+00.79 \citep{samal2018bipolar}, where  ionised lobes are separated by dense molecular gas. Using \textit{Herschel} and \textit{Spitzer} observations, \citet{deharveng2015bipolar, samal2018bipolar} show that many bipolar nebulae are associated with massive stars embedded within dense filaments or flattened molecular structures, supporting a direct link between bipolar morphology and radiative feedback within anisotropic density distributions. Numerical experiments further support this picture. \citet{Wareing2017} demonstrate that massive stars forming within dense sheet-like clouds can drive bipolar ionised bubbles surrounded by rings of swept-up material whose projected morphologies resemble observed infrared bubbles. Similar conclusions have been reached in studies of feedback in stratified clouds and filamentary environments \citep{Geen2015, Mackey2015}.

In a turbulent ISM, CCCs are expected to occur frequently, and may trigger massive star formation as suggested by several authors \cite[see][]{habe1992,fukui2014,balfour2017,fukui2021}. Numerical experiments show that CCCs produce dense \SCL s, thereby creating favourable conditions for rapid fragmentation and massive core formation \citep{inoue2013formation,takahira2014collisions}. The hydrodynamic experiments described by \citet{balfour2015star} demonstrate that for low-velocity CCCs the \SCL\ fragments to produce a \HFS, leading to the formation of a monolithic star cluster, containing massive stars. \citet{georgatos} show that the inclusion of a magnetic field in models of CCCs promotes the formation of massive stars, by significantly increasing the velocity threshold below which \HFS s are formed.\footnote{There are alternative mechanisms proposed for the formation of \HFS s. \citet{myers2009} points out that an isolated non-spherical cloud may collapse to a pancake, and this in turn may lead to the formation of a \HFS. This is also the setup invoked in the MHD simulations of \HFS\ formation reported by \citet{SuinPetal2025AA698A119}. \citet{kumar2020} develops a third scenario in which two pre-existing filaments collide. We discuss the merits of these alternative mechanisms, in the context of the formation of Bipolar \HIIR s, in Section \ref{SEC:BPHIIR.CCC}.} Moreover, the geometry of the \SCL\ means that ionising radiation readily escapes in directions orthogonal to the layer (i.e. parallel and anti-parallel to the collision axis), but is confined in directions perpendicular to the collision axis, thereby producing a Bipolar \HIIR\ \citep{whitworth2018bipolar} with a dense ring and Bright Rim at its waist.

In this paper we use three-dimensional hydrodynamic models to explore the formation, morphology, and lifetime of Bipolar \HIIR s formed by CCCs, focussing on the effects of different collision velocities and different levels of turbulence in the pre-collision clouds. Additionally, we investigate how viewing angle affects the appearance of these systems in ionised-gas and dust-continuum emission maps, with the goal of identifying metrics that might be used to identify Bipolar \HIIR s. The paper is organised as follows. Section \ref{sec:numerical} describes the numerical methods used. Section \ref{sec:results} presents synthetic observations of the results. Section \ref{sec:Disc} discusses the results, and Section \ref{SEC:Conc} summarises the main conclusions.

We stress that our models are not simulations, but rather experiments, designed to establish proof-of-concept, with very limited and simplified initial conditions -- rather than to reproduce observed systems. Therefore we have limited comparison with observations to generic, qualitative metrics. In a future paper we will expand the parameter space of the initial setups (both the number of parameters and their ranges), and make more quantitative comparisons with observations.

%%%%%
\section{Numerical Methods}\label{sec:numerical}
%%%%%

We perform three-dimensional radiation-hydrodynamic modelling of CCCs, using the adaptive mesh refinement code \textsc{flash} \citep{fryxell2000flash}. The computational domain is a cubic box with dimension $[8\,{\rm pc}]^3$, aligned with the Cartesian axes $[x,y,z]$. The base grid resolution generates $256^3$ cells, (yielding $0.031\,{\rm pc}$ resolution), and we impose an additional level of refinement within a central sphere of radius $1\,{\rm pc}$ (yielding $0.016\,{\rm pc}$ resolution).

The initial conditions involve two clouds, each with radius $\RO\seq2\pc$, mass $\MO\seq500\Msun$, and density $\rhoO\seq10^{-21}\gccc$. The clouds are placed at $[x,y,z]\seq[\pm2,0,0]$, with velocities $\mathbf{u}\seq[\mp\uO,0,0]$  so that they collide head-on immediately. In the sequel, the $x$ axis is termed `the collision axis', and viewing directions are specified using the angle, $\theta$, between the observer's line of sight and the collision axis. Thus the limiting angles are $\theta\seq0^\circ$ (looking along the collision axis and face-on to the \SCL) and $\theta\seq90^\circ$ (looking orthogonal to the collision axis and edge-on to the \SCL). The observers sky is represented by upper-case coordinates $[X,Y]$ (to distinguish them from the lower-case coordinates  $[x,y,z]$, of the computational domain).

We consider two collision velocities, $2\uO\seq2.4\,{\rm km\,s^{-1}}$ (hereafter the \lv\ collision) and $2\uO\seq4.0\,{\rm km\,s^{-1}}$ (hereafter the \hv\ collision). These collision velocities are informed by the assumption that the clouds are part of a fractal hierarchy, and therefore their bulk velocities reflect the velocity dispersion within a larger cloud-complex of $2,000\;\mbox{to}\;20,000\Msun$ that subscribes to Larson's scaling relations \citep{Larson1981}.  

Each cloud has an internal turbulent velocity field, characterised by a thermal mix of solenoidal and compressive modes, and a mean Mach Number, $\Mbar$. We treat $\Mbar\seq1$ (sonic turbulence), $\Mbar\seq3$ (supersonic turbulence), and $\Mbar\seq6$ (hypersonic turbulence). This allows us to evaluate the consequences of different levels of pre-existing internal substructure in the colliding clouds.

The chemical and thermal evolution is tracked using a network that computes the abundances of five species (H$_2$, H$^{\rm o}$, H$^+$, CO, CO$^+$), heating by compression and cosmic rays, cooling by line emission, and energy exchange with dust \citep{Federrath2010turb, Walch2015, wunsch2018}. Initial abundances are $\nHM\seq96\ccc$, $\nHO\seq226\ccc$, $\nHp\seq0$, $\nCO\seq0.5\ccc$ and $\nCOp\seq0$. The initial temperature in the clouds is $\TO\seq10\K$. The clouds are embedded in a fully ionised, low-density, hot background medium with density $\rho_{\mathrm{back}}\seq10^{-25}\gccc$ and temperature $T_{\rm back}=10^5\K$, in order to ensure approximate pressure balance across the cloud boundaries.

Sink particles are introduced according to the \cite{Federrath2010} algorithm, with $\rhosink\seq10^{-19}\gccc$ and $\Rsink\seq0.045\pc$ (i.e. $\sim\!3$ grid-cells in the central region), thereby ensuring that the Truelove condition \citep{Truelove1997} is satisfied. Since individual stars are not resolved (individual sinks start with $\sim\!0.5\Msun$ and usually grow a lot larger), we model feedback using the \textsc{feedbacksinks} module  \citep{GattoAetal2017MN466p1903, Walch2015, Haid2019}. \textsc{feedbacksinks} assigns a random stellar population to each sink, based on its total mass and a prescribed initial mass function, thereby allowing the ionising luminosity to be estimated self-consistently. An OB star is only formed once the sink mass exceeds a threshold of $120\Msun$. To allow for the stochastic nature of the algorithm, we run two realisations of each setup, using two different random seeds. 

Ionising radiation is treated using the \textsc{treeray} radiation transport module \citep{wunsch2021tree}, which estimates the radiation field by casting rays from each grid cell, using a tree-based angular decomposition. This enables nearby sources to be treated individually, whilst distant sources are grouped into a single effective emitter. This provides an efficient estimate of the ionising radiation field in clustered star-forming environments. We exclude the effects of radiation pressure, stellar winds and supernovae. These effects are also likely to play an important role  \citep{krumholz2014role, Dale2015, heyer2015phases, kim2017feedback, hopkins2018feedback, geen2017impact}, and will be included in future papers, but here we seek to isolate the effects of ionising feedback.

Our experiments explore a total of 12 setups: two random realisations each, for all possible combinations of two collisions velocities ($2\uO$) and three turbulent Mach Numbers ($\Mbar$). The parameters defining the initial conditions for the experiments are summarised in Table \ref{table 1}. In addition, we have modelled the evolution of a single cloud, with hypersonic internal turbulence ($\Mbar\seq6$) and a sink at its centre having initial mass $M\!\sim\!120\Msun$ (i.e. just at the threshold to form an OB star according to the \citet{GattoAetal2017MN466p1903} subgrid model), in order to obtain the solution for an \HIIR\ expanding approximately isotropically. This setup is referred to as the `Fiducial Setup', and is used for comparison with the CCC Setups.

There is an important distinction between the geometry of the CCC Setups, and the geometry of the Fiducial Setup. Without the turbulence imposed on the pre-collision clouds, the CCC Setups are cylindrically symmetric about the $x$ axis, whereas the Fiducial Setup is spherically symmetric about the origin.

%%%%%
\begin{table}
\caption{Parameters describing the initial conditions.}
\begin{center}
\begin{tabular}{ll}\hline
Initial cloud radius & $R_{\rm o} = 2$\,pc \\
Cloud mass & $M_{\rm o} = 500\, M_{\odot}$\\
Initial cloud density & $\rho_{\rm o} \simeq 10^{-21}\gccc$ \\
Background density & $\rho_{\rm back} = 10^{-25}\gccc$ \\
Relative collision velocities & $\Delta u_{\rm o}=2u_{\rm o} = 2.4 \textrm{ and }
4.0\kms$ \\
Turbulence Mach number & $\Mbar = 1,\;3\;\,\textrm{and}\,\;6$ \\
Sink creation density threshold & $\rhosink = 10^{-19}$\,g\,cm$^{-3}$ \\
Sink radius & $\Rsink = 0.045\,{\rm pc}$ \\
Size of the computational domain & $L_{\rm o} = 8$\,pc \\\hline
\end{tabular}
\end{center}
\label{table 1}
\end{table}
%%%%%

%%%%%
\section{Results}\label{sec:results}
%%%%%

%%%%%
\subsection{Star formation in the \SCL\ }
%%%%%

%%%%%

\begin{figure*}
\vspace{-0.2cm}
\centering
\includegraphics[width=1\linewidth]{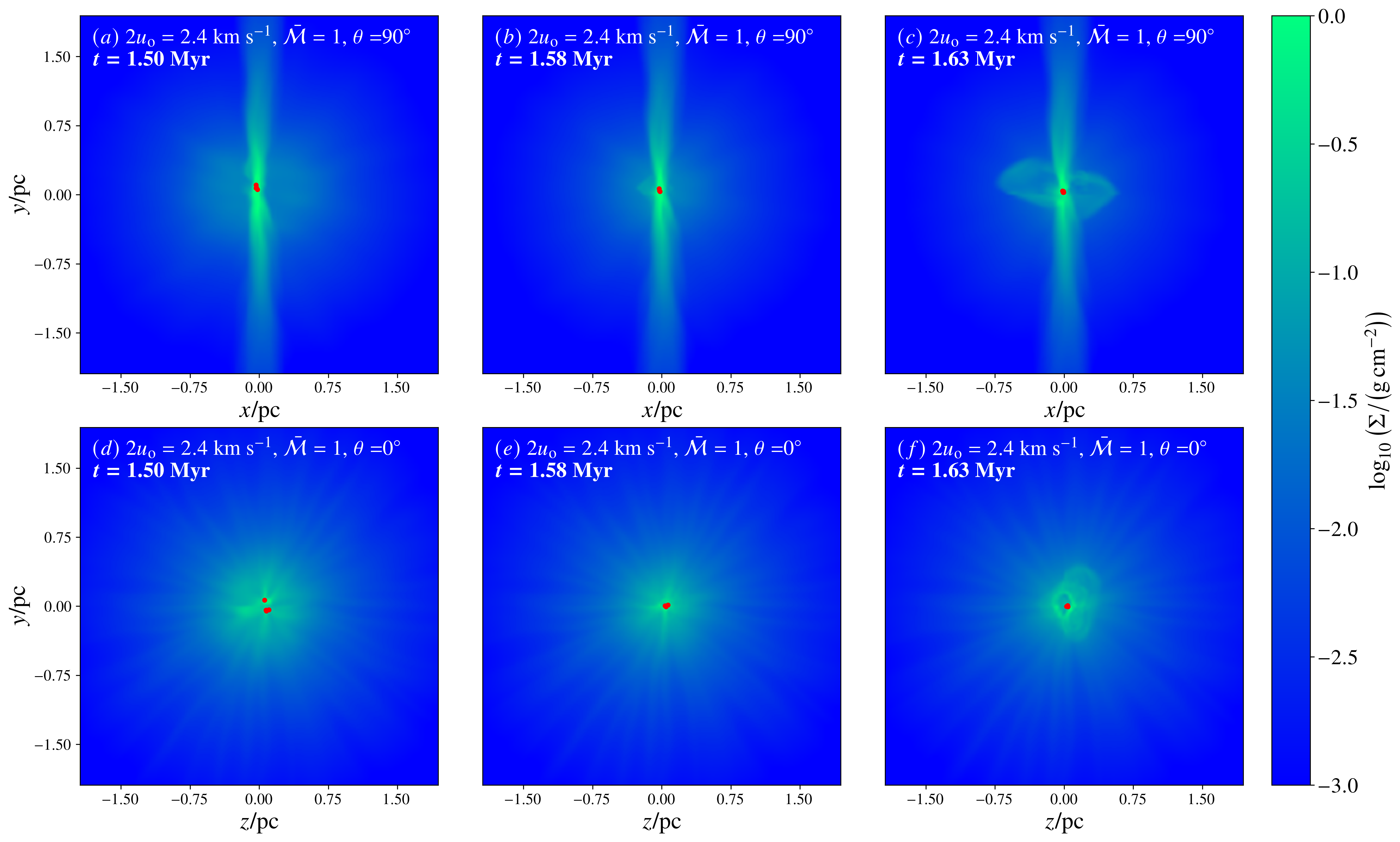}\vspace{0.4cm}
\includegraphics[width=1\linewidth]{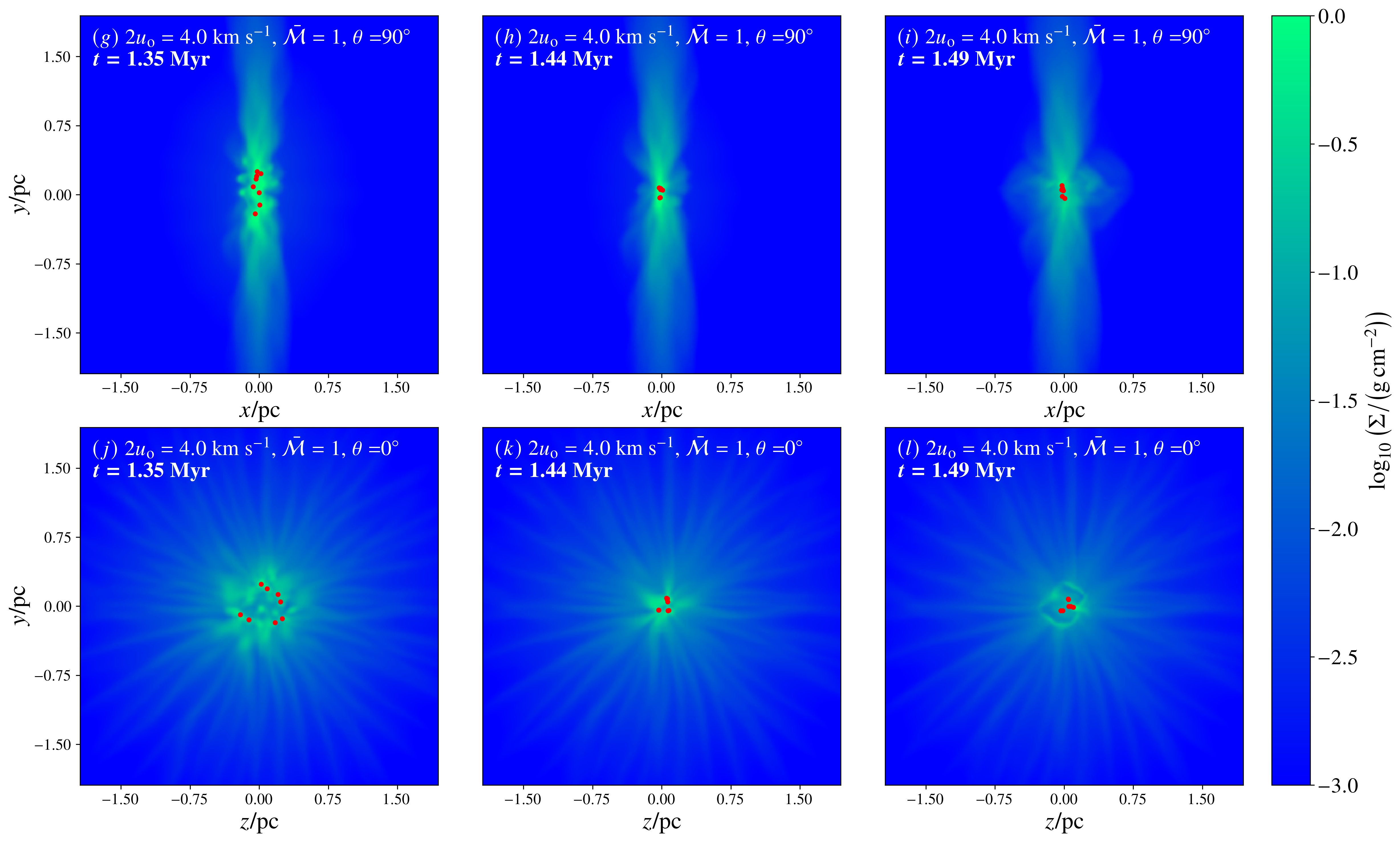}
\vspace{-0.6cm}
\caption{False-colour column-density maps for two CCCs, both with sonic internal turbulence ($\mathcal{M}\seq1$). The top six panels {\it (a} to {\it f)} represent a low-velocity collision ($2\uO\seq2.4\kms$), and the bottom six panels ({\it g} to {\it l}) represent a high-velocity collision ($2\uO\seq4.0\kms$). Within each velocity group, the upper row (rows 1 and 3; panels {\it (a} to {\it c)}, and {\it (g} to {\it i)}) shows the projection onto the $xy$-plane (i.e. looking at the Shock-Compressed Layer edge-on, viewing angle $\theta\seq90^\circ$), and the lower row (rows 2 and 4; panels {\it (d} to {\it f)}, and {\it (j} to {\it l)}) shows the projection onto the $yz$ plane (i.e. looking at the Shock-Compressed Layer face-on, viewing angle $\theta\seq0^\circ$). From left to right, the columns represent the time when $10\%$ of the total cloud mass has been assimilated by sinks (lefthand column); the time when the first OB star is formed (middle column); and the time $50\kyr$ after the formation of the first OB-type star (righthand column). The values of $2\uO$, $\Mbar$, $t$ and $\theta$ are given at the top of each panel. The column-density scale is logarithmic. Red dots mark the positions of sink particles.} 
\label{fig:ColDen}
\end{figure*}
%%%%%

Using SPH experiments, \citet{balfour2015star} and \citet{georgatos} identify two distinct fragmentation modes for \SCL s produced by CCCs. Following a \lv\ collision the layer fragments into a \HFS, and most of the star formation occurs in the Hub, producing a monolithic star cluster with some very massive stars. In contrast, following a \hv\ collision the layer initially fragments into a network of filaments (a \SWS) with small subclusters condensing out at the intersections of the filaments and no very massive stars. These experiments did not include feedback, and were terminated once $\sim\!10\%$ of the cloud mass had been converted into sinks. By implementing the \textsc{treeray} module in \textsc{flash}, we are able to model radiative feedback from sinks and follow the evolution until the \SCL\ is disrupted by ionising radiation.

Figure \ref{fig:ColDen} shows column-density maps from a \lv\ collision ($2\uO\seq2.4\kms$; top six panels, {\it (a)} to {\it (f)}) and a \hv\ collision ($2\uO\seq4.0\kms$; bottom six panels, {\it (g)} to {\it (l)}). The \SCL\ is viewed edge-on on the first and third rows (panels {\it (a)} to {\it (c)} and {\it (g)} to {\it  (i)}), and face-on on the second and fourth rows (panels {\it (d)} to {\it (f)} and {\it (j)} to {\it (l)}). On each row the three maps show a time-sequence, with time increasing from left to right. In both setups the pre-collision clouds have sonic turbulence ($\Mbar\seq1$), and peak column-densities reach $\sim\!1\gcc$ (equivalently, $\NHM\sim\!2\times 10^{23}\cc$). See figure caption for further details.

%%%%%
\subsubsection{Low-velocity collisions}
%%%%%

Low-velocity collisions with sonic turbulence produce relatively thin \SCL s that fragment to produce \HFS s. Sink particles form primarily in the central Hub, where they grow in mass until the threshold for OB star formation is passed and they start emitting ionising radiation. The ionised gas expands and the \IF\ quickly breaks out of the layer in directions close to the collision axis (the $x$ axis), producing the two lobes of a Bipolar \HIIR. In directions orthogonal to the collision axis (close to the $yz$ plane), expansion of the \IF\ is slow, because it runs into the dense \SCL, and a dense ring is swept up, bounded by a \BR. This \BR\ defines the waist of the Bipolar \HIIR.

Viewed orthogonal to the collision axis ($\theta\seq90^\circ$; panels {\it (a)} to {\it (c)} on the first row of Figure \ref{fig:ColDen}) the \SCL\ appears as a narrow column-density maximum parallel to the $y$ axis. The bipolar lobes are visible on the final map. Viewed along the collision axis ($\theta\seq0^\circ$; panels {\it (d)} to {\it (f)} on the second row of Figure \ref{fig:ColDen}) the filaments feeding the central Hub can be seen at early times, and the dense ring defining the waist of the Bipolar \HIIR\ is just visible on the final map. We note that at this stage a substantial part of each cloud is still flowing towards the \SCL.

%%%%%
\subsubsection{High-velocity collisions}
%%%%%

The \SCL s produced by \hv\ collisions initially fragment into a network of filaments (a \SWS), with sinks forming preferentially at the intersections of filaments. Sinks form earlier, and with larger separations than in \lv\ collisions (because the surface-density of the \SCL\ builds up faster and there is less time for the \SCL\ to contract towards the $x$ axis). However, the Spiders-Web phase is transient. Both gas and sinks fall towards a central Hub, and OB stars only form after the Hub has been assembled. No OB stars are formed during the Spiders-Web phase.

The structure of the \SCL\ differs markedly from the \lv\ setup. Vishniac instabilities produce a thicker, less coherent \SCL, as shown on the third row of Figure \ref{fig:ColDen} (panels {\it (g)} to {\it (i)}). Consequently the waist of the Bipolar \HIIR\ expands faster (panel {\it (l)}, on the fourth row of Figure \ref{fig:ColDen}). Sinks are initially formed quite far apart (see panel {\it (j)} on the fourth row of Figure \ref{fig:ColDen}), but by the time they start emitting ionising radiation (panel {\it (k)} on the fourth row of Figure \ref{fig:ColDen}) they are in a compact central cluster.

%%%%%
\subsubsection{Influence of turbulence}
%%%%%

%%%%%
\begin{figure*}
\centering
\includegraphics[width=1\linewidth]{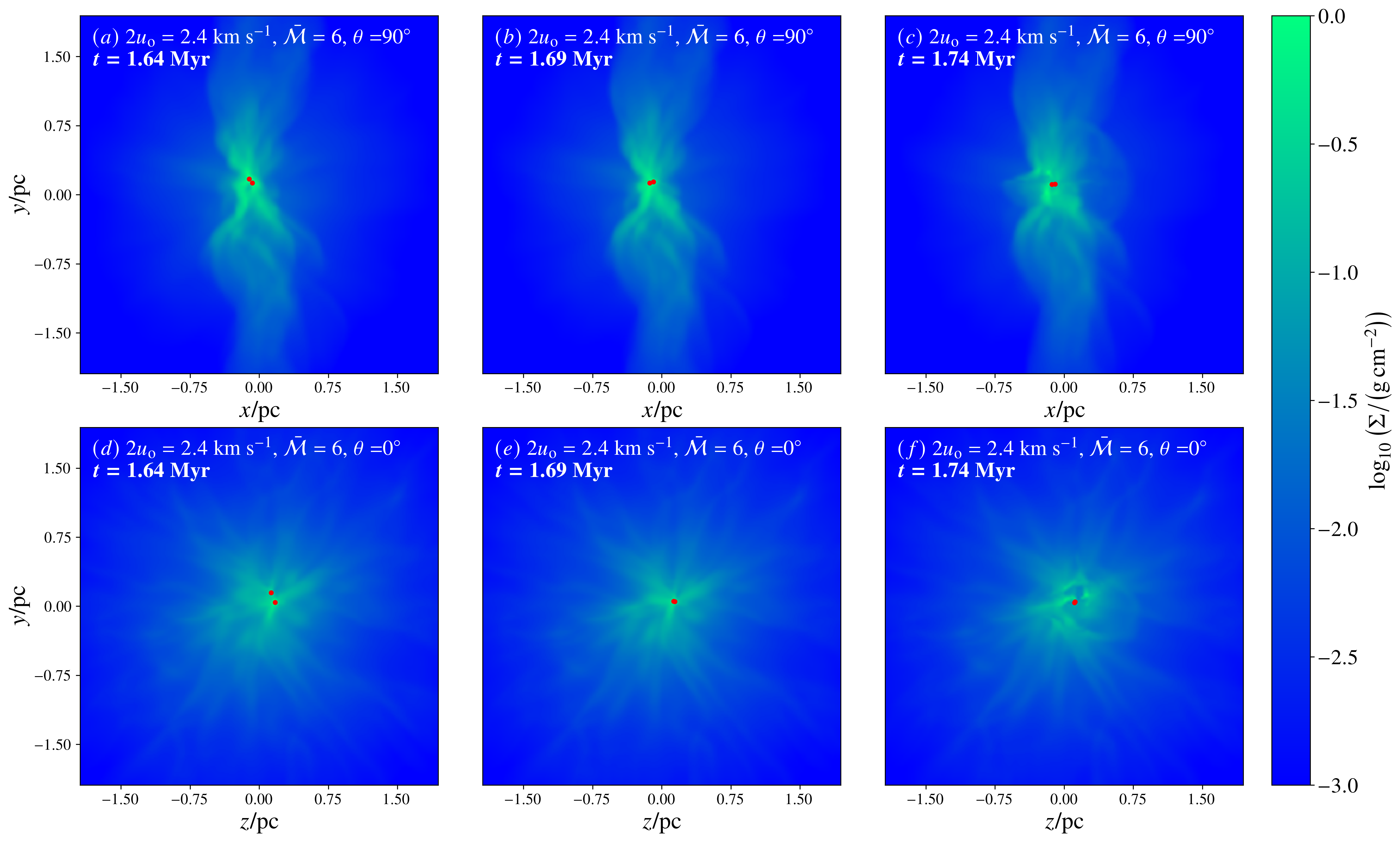}\vspace{0.5cm}
\includegraphics[width=1\linewidth]{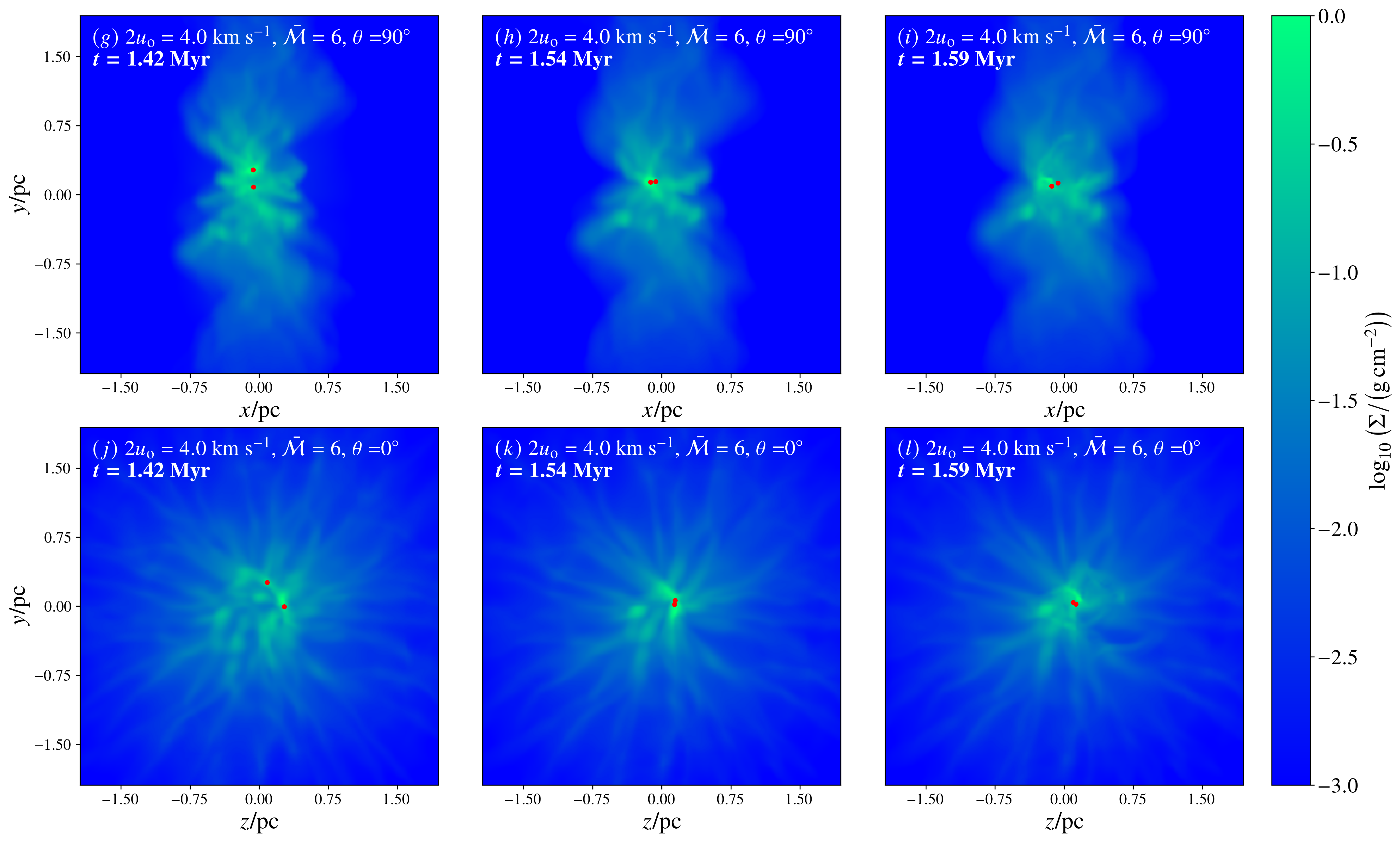}
\vspace{-0.2cm}
\caption{As Figure \ref{fig:ColDen} but for clouds with hypersonic internal turbulence ($\Mbar\seq6$).}
\label{fig:turbulence}
\end{figure*}
%%%%%

The effect of increasing the level of turbulence in the pre-collision clouds, from sonic ($\Mbar\seq1$) to hypersonic ($\Mbar\seq6$), is illustrated on Fig. \ref{fig:turbulence}. The panels on Figure \ref{fig:turbulence} are arranged as on Figure \ref{fig:ColDen}, as far as collision velocity, viewing angle and time are concerned; only the level of turbulence is increased. Increasing the turbulent Mach Number does not significantly alter the overall fragmentation mode of the \SCL. \Lv\ collisions fragment into \HFS s, whilst \hv\ collisions initially produce a \SWS, which then collapses into a central cluster.

However, increasing the turbulent Mach number, and hence turbulent support against compression, does delay star formation somewhat. In the setup with hypersonic turbulence ($\Mbar\seq6$), both the time at which $10\%$ of the gas mass has been converted into sinks, and the formation time of the first OB star, are extended by $\sim\!10\%$. This in turn means that almost all the mass of the colliding clouds is in the SCL by the time OB stars form.

Increased turbulence also modifies the morphology of the \SCL. In the setups with hypersonic turbulence, the \SCL\ is initially much broader and less coherent than in the setups with sonic turbulence. This is because there is more pronounced substructure in the pre-collision gas. Combined with the delayed onset of ionising feedback, this has two consequences. Firstly, there is little gas left at large $|x|$ that can be ionised to produce bipolar lobes; this is shown on panel {\it (i)} of Figure \ref{fig:turbulence}. Secondly, the boundary where the ionising radiation meets the dense infalling gas of the \SCL\ is more irregular, and so the waist is also irregular, and does not resemble a ring, as shown on panel {\it (l)} of Figure \ref{fig:turbulence}.

%%%%%
\begin{figure}
\centering
\includegraphics[width=1\linewidth]{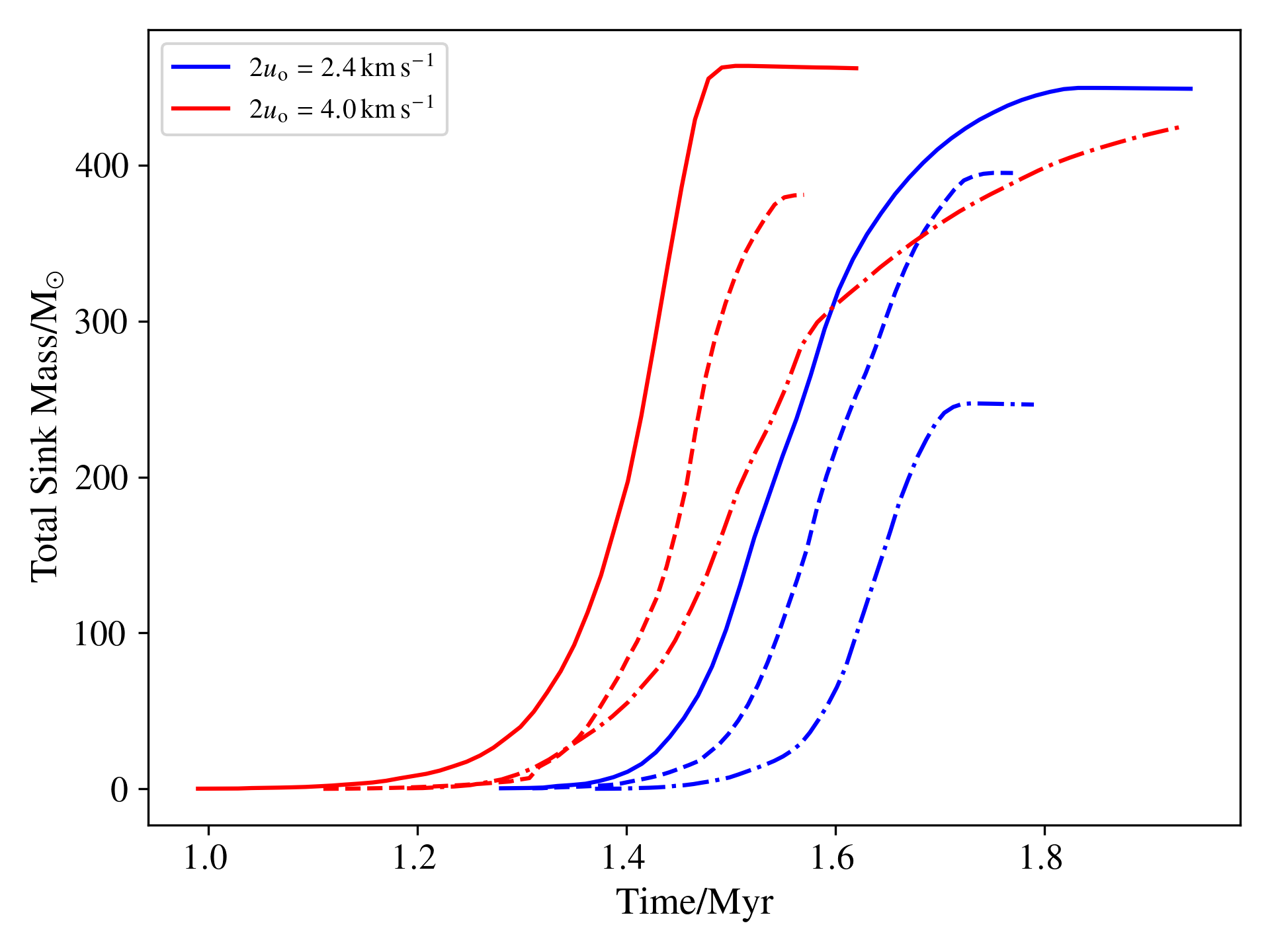}
\vspace{-0.4cm}
\caption{The total mass in stars as a function of time for all setups. Blue curves represent \lv\ collisions ($2\uO\seq2.4\kms$), and red curves represent \hv\ collisions ($2\uO\seq4.0\kms$). Full lines are for clouds with sonic turbulence ($\Mbar\seq1$), dash-dotted lines are for clouds with supersonic turbulence ($\Mbar\seq3$), and dashed lines are for clouds with hypersonic turbulence ($\Mbar\seq6$). Each line gives the mean of two random realisations.}
\label{fig:SFE}
\end{figure}
%%%%%

%%%%%
\subsubsection{The Star Formation Rate}
%%%%%

Figure \ref{fig:SFE} shows the total mass in stars, as a function of time, for different collision velocities, and for different levels of turbulence in the pre-collision clouds. Once sufficient material has been assembled into a central Hub, the star formation rate becomes quite large, reaching values between $\sim\!2.5\Msun\kyr^{-1}$ and $\sim\!4.0\Msun\kyr^{-1}$. In general, star formation starts earlier, proceeds more rapidly, and terminates more abruptly when the clouds collide at high velocity and the turbulence level is low. With high collision velocity, the \SCL\  is assembled faster. And with lower turbulence, there is then less resistance to gravitational compression. In all cases star formation is quite rapid; most of the sink mass is assimilated in a short burst, $\lesssim\!0.1\Myr$, between when sufficient dense gas has been assembled in the Hub, and when ionising feedback has dispersed the residual gas (which is typically $65\%$ of the total).

%%%%%
\begin{figure*}
\centering
\includegraphics[width=1\linewidth]{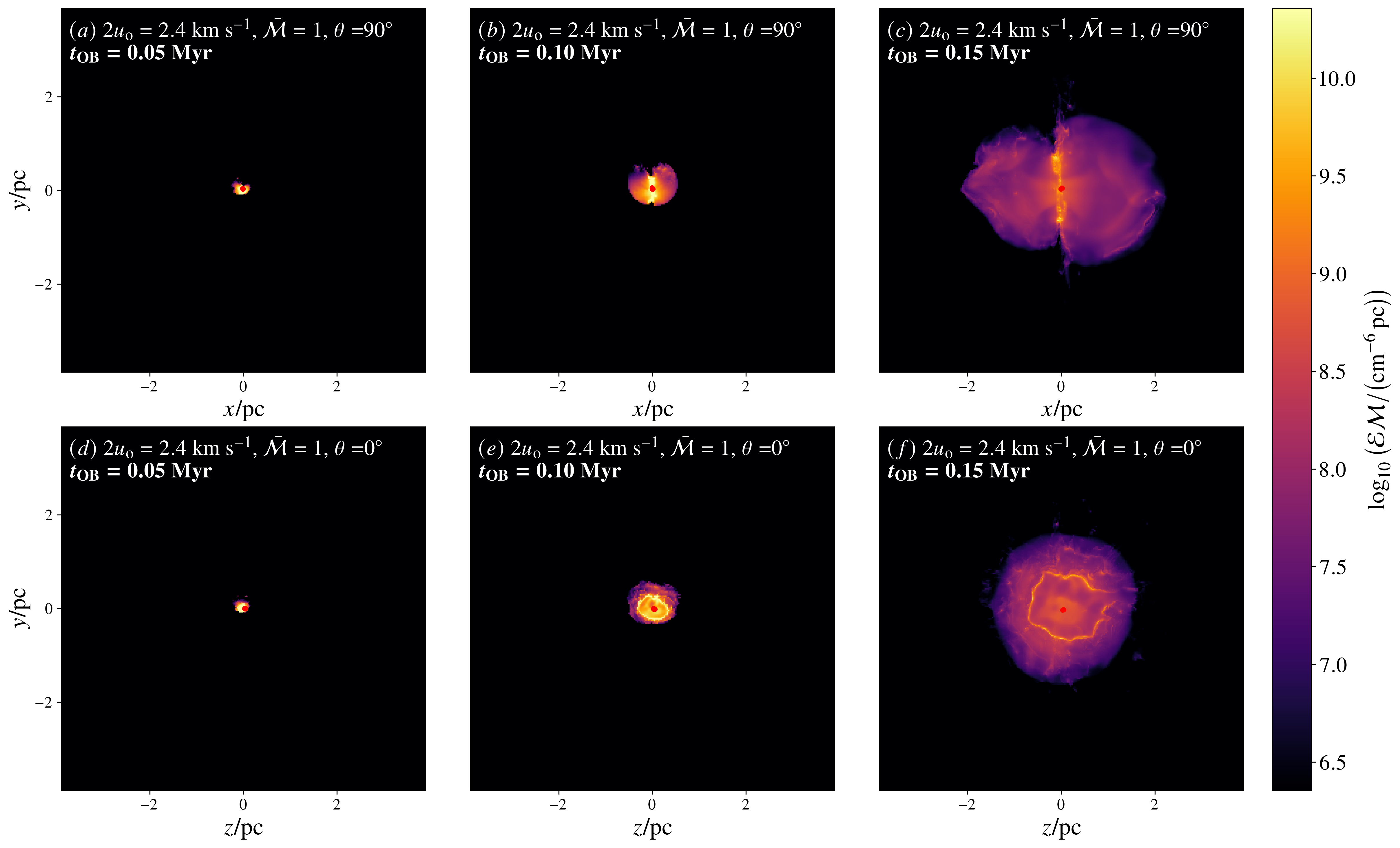}\vspace{0.4cm}
\includegraphics[width=1\linewidth]{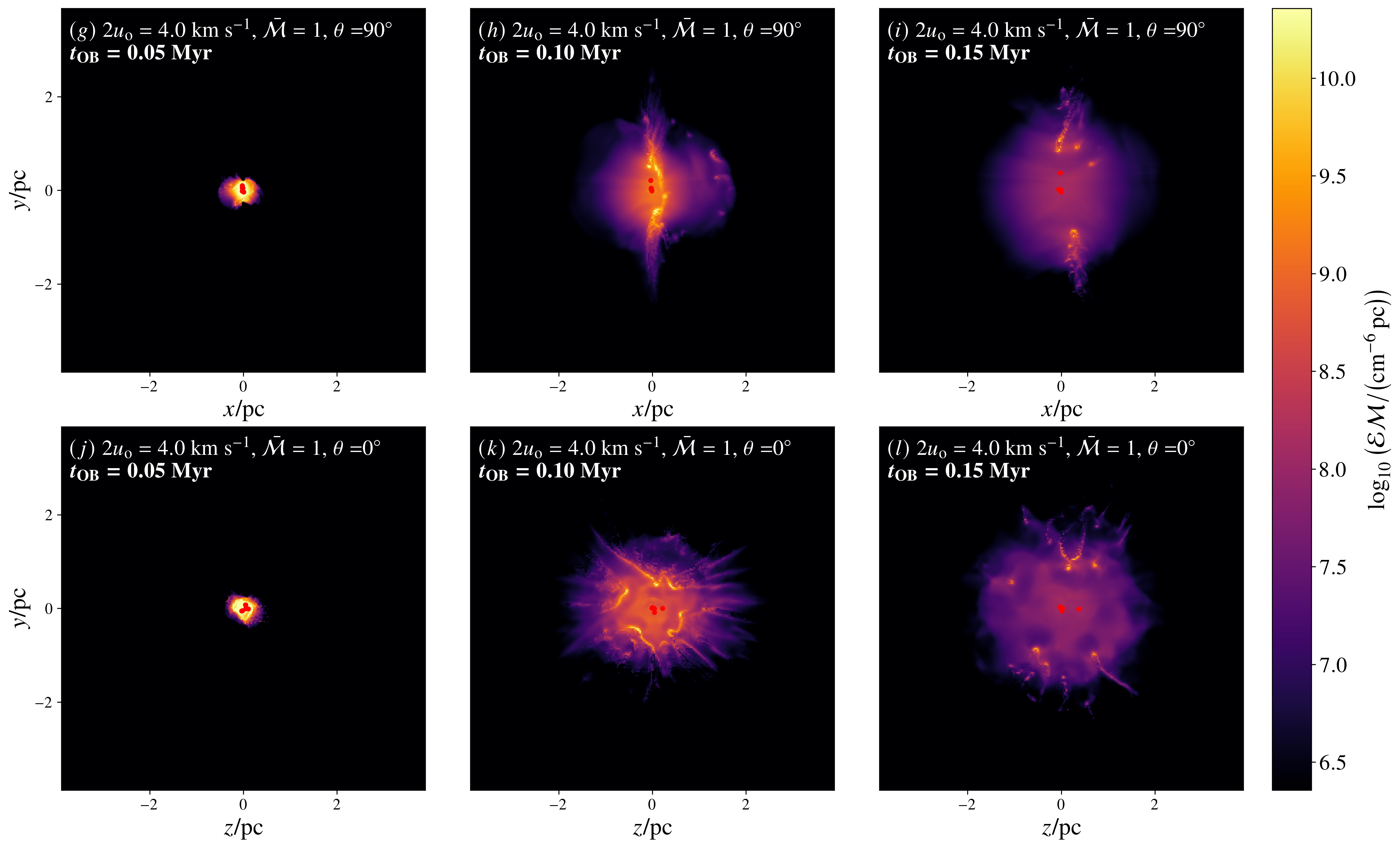}
\vspace{-0.6cm}
\caption{False-colour maps of the emission measure for setups with sonic turbulence ($\Mbar\seq1$). The top six panels {\it ((a)} to {\it (f)}) represent a \lv\ collision ($2\uO\seq2.4\kms$), and the bottom six panels {\it ((g)} to {\it (l)}) represent a \hv\ collision ($2\uO\seq4.0\kms$). Within each velocity group, the upper row (rows 1 and 3; panels {\it (a)} to {\it (c)}, and {\it (g)} to {\it (i)}) shows the projection onto the $xy$-plane (i.e. looking at the Shock-Compressed Layer edge-on, viewing angle $\theta\seq90^\circ$), and the lower row (rows 2 and 4; panels {\it (d)} to {\it (f)}, and {\it (j)} to {\it (l)}) shows the projection onto the $yz$ plane (i.e. looking at the Shock-Compressed Layer face-on, viewing angle $\theta\seq0^\circ$). From left to right, the three columns represent a time-sequence at $\tOB\seq0.05\Myr$, $\;0.10\Myr\,$ and $\;0.15\Myr\,$ after formation of the first OB-type star. The values of $2\uO$, $\Mbar$, $\tOB$ and $\theta$ are given at the top of each panel. As is standard practice, the Emission Measure is given in ${\rm cm^{-6}\,pc}$, and is scaled logarithmically. Red dots mark the locations of sink particles.}
\label{fig:emission2.4}
\end{figure*}
%%%%%

%%%%% 
\begin{figure}
\centering
\includegraphics[width=\linewidth]{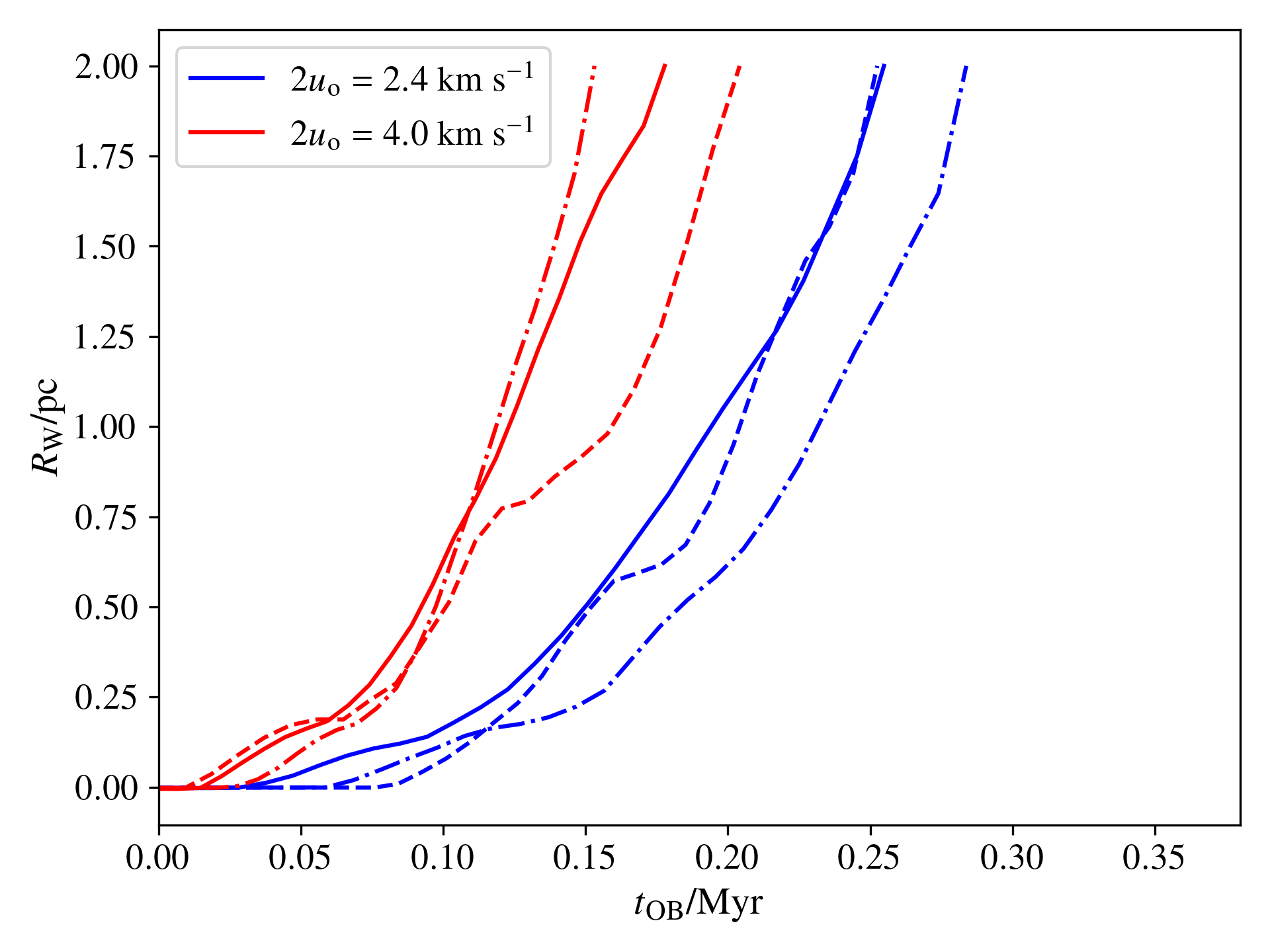}
\vspace{-0.5cm}
\caption{The waist radius ($\RW$) as a function of time since formation of the first OB star ($\tOB$). Red curves represent \hv\ collisions and blue curves represent \lv\ collisions. Solid, dash-dotted and dashed lines represent, respectively, setups with sonic ($\Mbar\seq1$), supersonic ($\Mbar\seq3$), and hypersonic ($\Mbar\seq6$) turbulence. Each curve is the mean of two realisations.}
\label{fig:waist}
\vspace{-0.4cm}
\end{figure}
%%%%%

%%%%%
\begin{figure*}
\includegraphics[width= \linewidth]{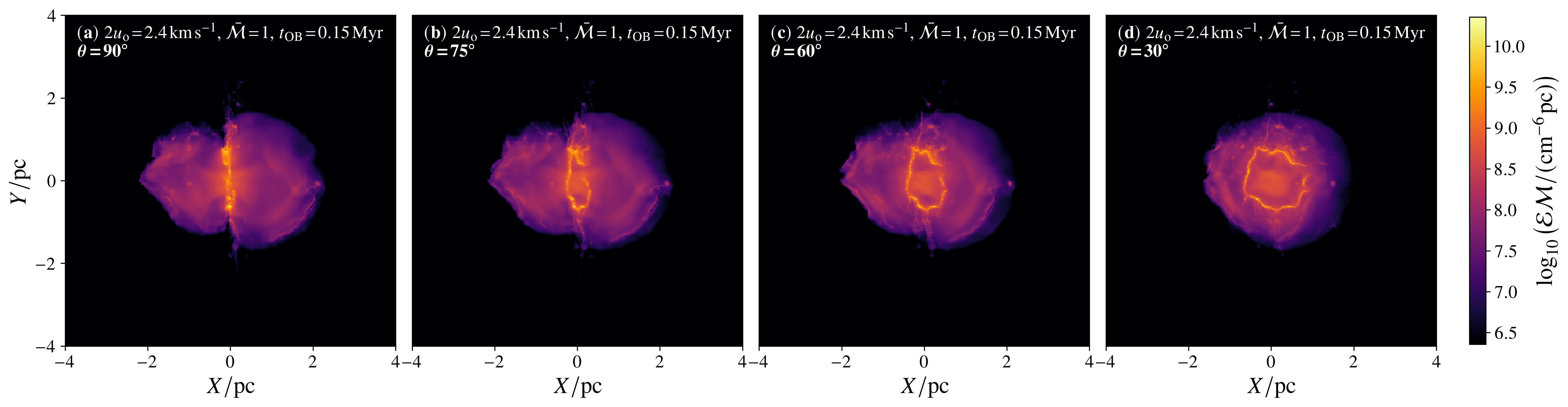}
\includegraphics[width = \linewidth]{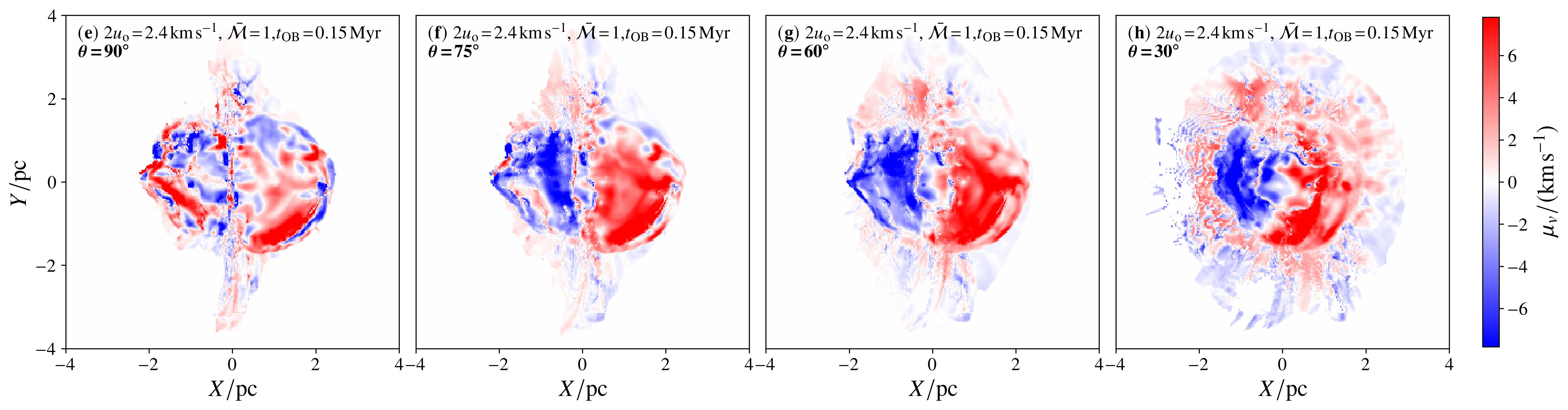}
\caption{{\it Top row}: false-colour maps of Emission-Measure for a \lv\ collision ($2\uO\seq2.4\kms$) with sonic turbulence in the pre-collision clouds ($\Mbar\seq1$), at $t_{\mathrm{OB}}=0.15$ Myr, from four different viewing angles: {\it (a)} $\theta\seq90^\circ$; {\it (b)} $\theta\seq75^\circ$; {\it (c)} $\theta\seq60^{\circ}$; {\it (d)} $\theta\seq30^\circ$. ~{\it Bottom row}: maps of the the corresponding Emission-Measure weighted mean line-of-sight velocity, $\muv$. The value of $\theta$ is given at the top of each panel. }
\label{fig:viewing}
\end{figure*}
%%%%%

%%%%%
\begin{figure*}
\includegraphics[width= \linewidth]{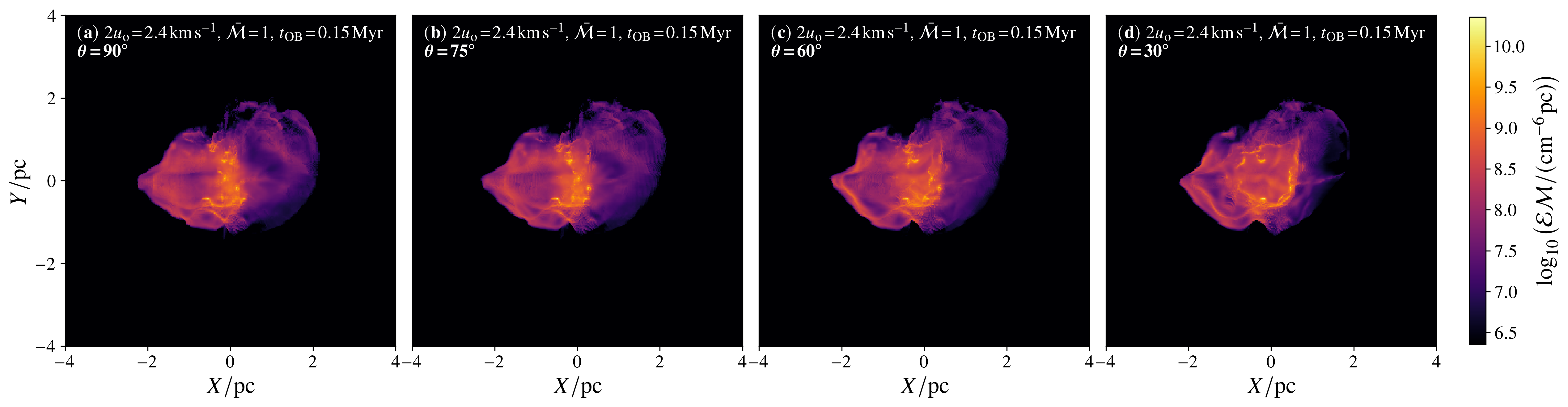}
\includegraphics[width = 1\linewidth]{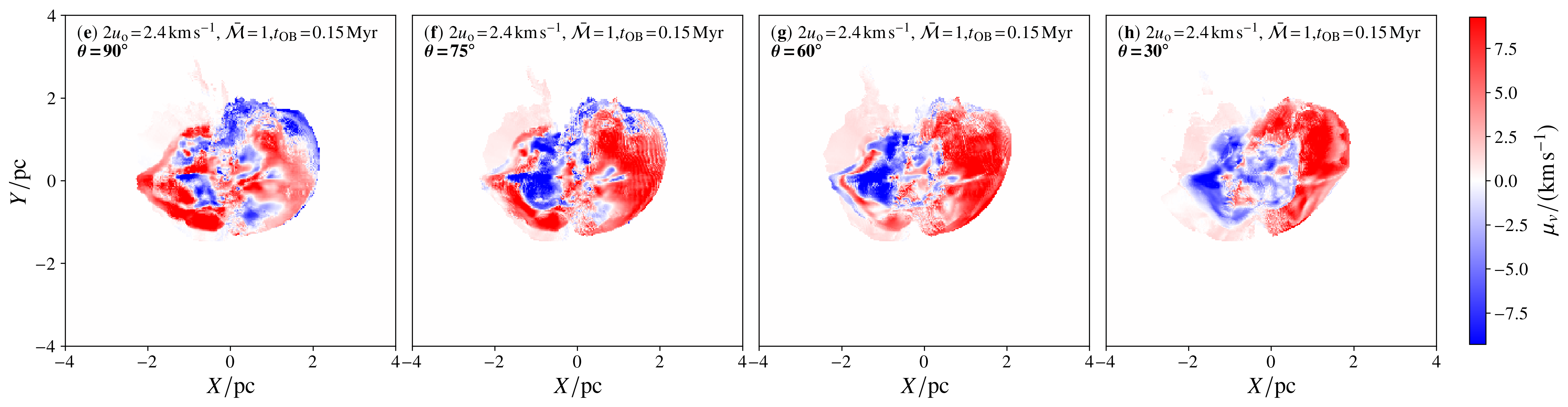}
\caption{As Figure \ref{fig:viewing}, but with hypersonic turbulence, $\Mbar\seq6$, in the pre-collision clouds (rather than sonic turbulence, $\Mbar\seq1$).}
\label{fig:viewingturb}
\end{figure*}
%%%%%

%%%%%
\begin{figure*}
\includegraphics[width= \linewidth]{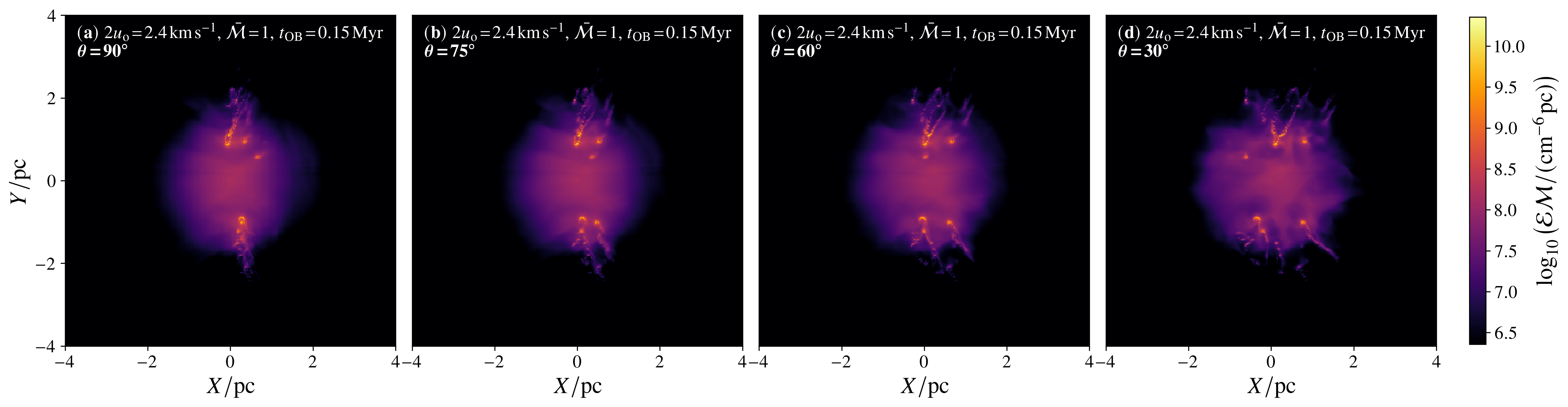}
\includegraphics[width = 1\linewidth]{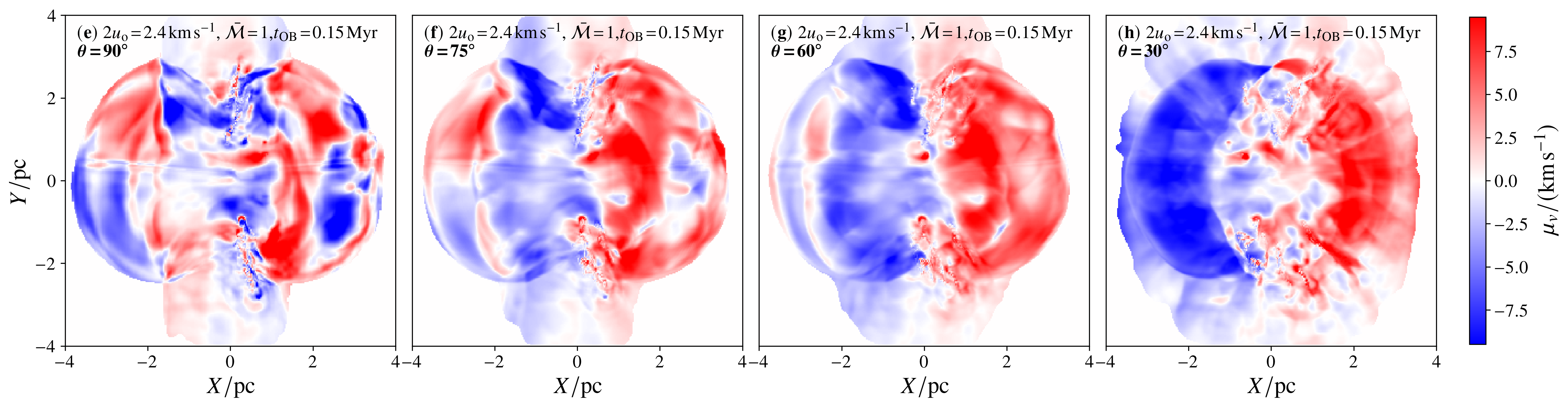}
\caption{As Figure \ref{fig:viewing}, but with higher collision velocity, $2\uO\seq4.0\kms$ (rather than $2\uO\seq2.4\kms$).}
\label{fig:viewing4_0}
\end{figure*}
%%%%%

%%%%%
\begin{figure*}
\includegraphics[width= \linewidth]{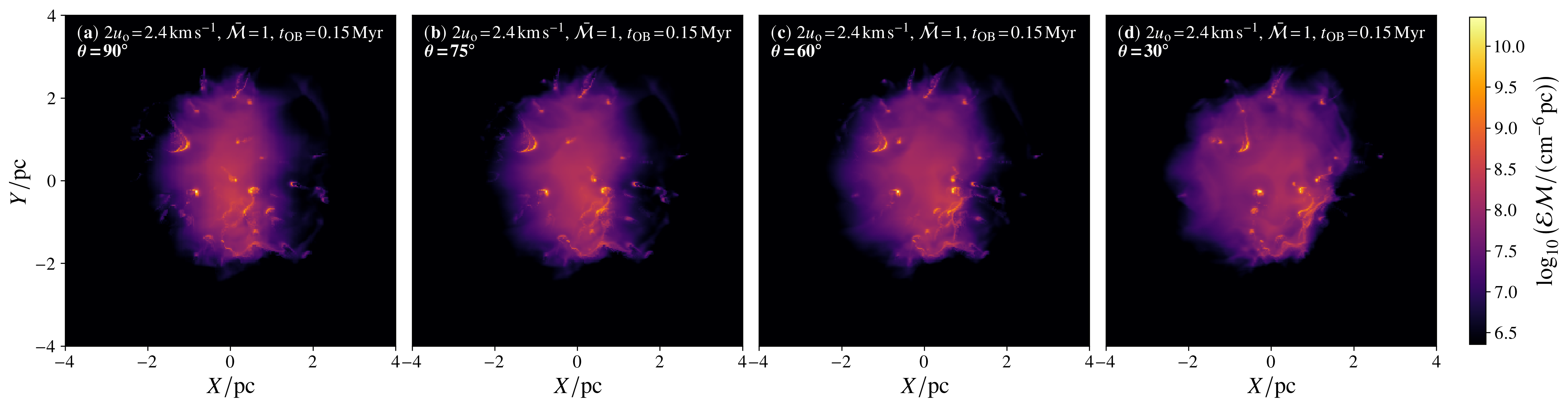}
\includegraphics[width = \linewidth]{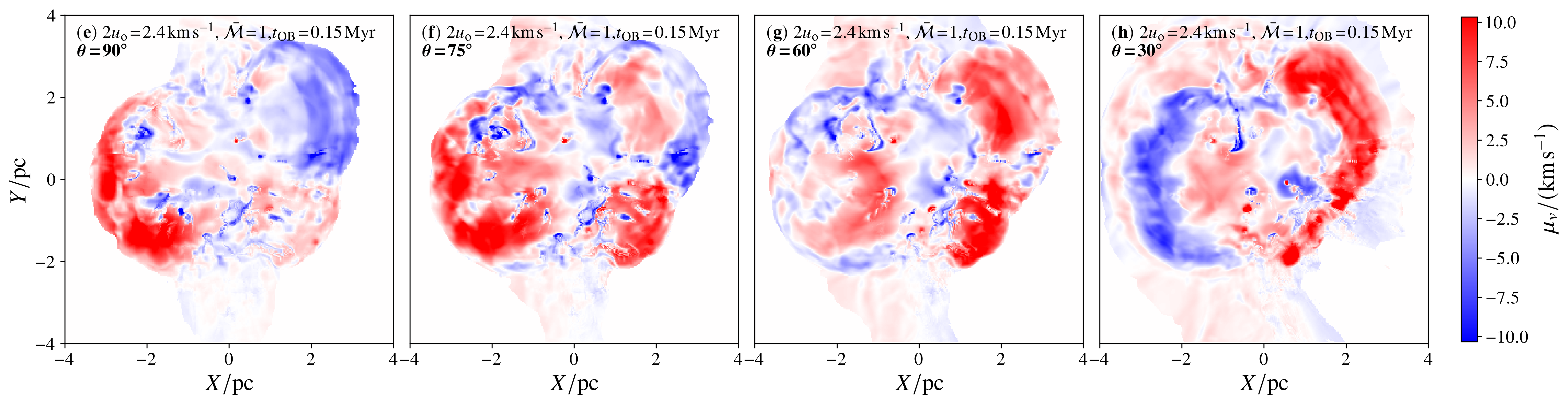}
\caption{As Figure \ref{fig:viewing}, but with higher collision velocity, $2\uO\seq4.0\kms$ (rather than $2\uO\seq2.4\kms$) and hypersonic turbulence, $\Mbar\seq6$, in the pre-collision clouds (rather than sonic turbulence, $\Mbar\seq1$).}
\label{fig:viewing4_0turb}
\end{figure*}
%%%%%

%%%%%
\begin{figure*}
\centering
\includegraphics[width = 1\linewidth]{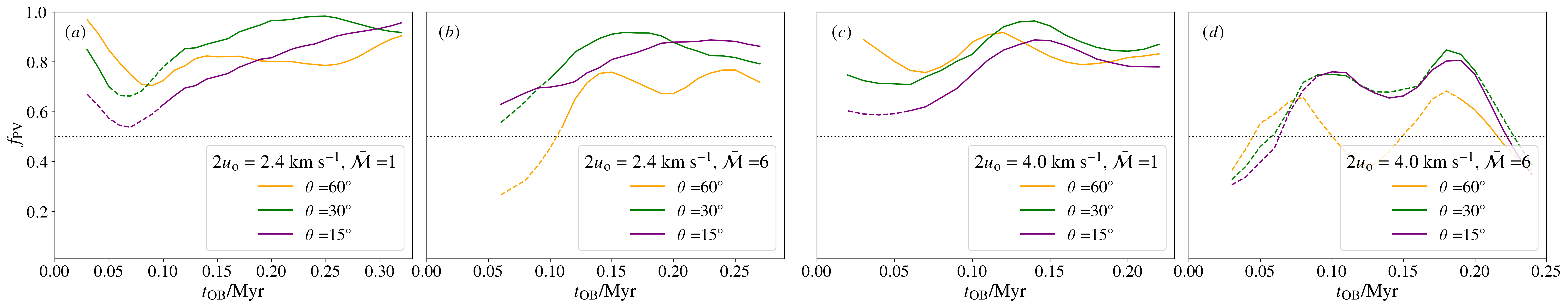}
\vspace{-0.3cm}
\caption{The fraction of pixels for which the sign of the mean Emission-Measure weighted radial velocity matches the sign of the sky coordinate $X$ (i.e. gas on the left side moving toward the observer, and gas on the right side moving away from the observer). {\it (a)} \lv\ collision with sonic turbulence; {\it (b)} \lv\ collision with hypersonic turbulence; {\it (c)} \hv\ collision with sonic turbulence; {\it (d)} \hv\ collision with hypersonic turbulence. The values of $2\uO$ and $\Mbar$ are given on each panel, along with the colour coded viewing angle: $\theta\seq60^\circ$ (green); $\theta\seq30^\circ$ (orange); $\theta\seq15^\circ$ (purple). Solid lines represent regions where the binomial test (against a null hypothesis of uniform velocity) is rejected with $p<0.05$, and dashed lines represent regions where $p>0.05$. Each curve is the mean of two realisations}
\label{fig:pv}
\end{figure*}
%%%%%

%%%%%
\subsection{Radio observations of the ionised gas}\label{SEC:RadioHII}
%%%%%

%%%%%
\subsubsection{Emission-Measure Maps}
%%%%%

\HIIR s can be observed in freefree continuum emission \citep[e.g.][]{luisi2016,delafuente2020}, and in recombination line emission \citep[e.g.][]{Churchwell2002}, as well as in various cooling lines from ionised metals (e.g. forbidden lines from O$^+$ and O$^{++}$). The intensities of freefree-continuum and recombination-line emission are determined by the Emission Measure along the line of sight,
\begin{eqnarray}\label{EQN:EmissionMeasure.01}
{\cal EM}&\simeq&\int\limits_{s=0}^{s=\infty}\;\nel\!(s)\;\npr\!(s)\;ds\,.\hspace{0.5cm}
\end{eqnarray}
Here $s$ is distance along the line of sight, $\nel\!(s)$ is the local electron number-density, $\npr\!(s)$ is the local proton number-density.

Under ideal conditions the observed intensities of freefree-continuum and recombination-line emission are exactly proportional to the Emission Measure. Specifically, the density ~{\it and} ~temperature in the \HIIR\ must be uniform, and the wavelength must be long, so that dust obscuration is negligible; additionally a correction for non-LTE effects must be included if the wavelength and/or density are low \citep[e.g.][]{peters2012}. If self absorption is important the intensity is not exactly proportional to the emission measure, but the relationship is straightforward.

The generation of maps of freefree-continuum intensity, or recombination-line intensity, in specific wavelength bands, lies outside the scope of this paper. Therefore we simply present maps of the Emission Measure, obtained by performing the integration of Equation \ref{EQN:EmissionMeasure.01} numerically, through the computational domain. These maps give a good indication of where the emission from the ionised gas is most intense, and how it is distributed on the observer's sky. The generation of maps of freefree-continuum and recombination line intensity will be treated in a future paper.

Because ionising radiation becomes dynamically relevant only after the formation of the first OB star, we define a new time variable $\tOB$, measured from the timestep when the first OB star forms.

%%%%%
\subsubsection{Morphological evolution of the H\,\textsc{ii} regions}
%%%%%

Figure \ref{fig:emission2.4} shows Emission-Measure maps for CCCs involving clouds with sonic turbulence ($\Mbar\seq1$). The top six panels {\it ((a)} to {\it (f)}) are for a \lv\ collision ($2\uO\seq2.4\kms$), and the bottom six panels {\it ((g)} to {\it (l)}) are for a \hv\ collision ($2\uO\seq4.0\kms$). Within each velocity group, the upper row (rows 1 and 3; panels {\it (a)} to {\it (c)}, and {\it (g)} to {\it (i)}) shows the \SCL\ edge-on (viewing angle $\theta\seq90^\circ$) and the lower row (rows 2 and 4; panels {\it (d)} to {\it (f)} and {\it (j)} to {\it (l)}) shows the \SCL\ face-on (viewing angle $\theta\seq0^\circ$). From left to right the different columns represent a time-sequence, and are labelled with $\tOB$.  

Following the \lv\ collision, two rather well-defined bipolar lobes develop, as seen in the edge-on view at $\tOB\seq0.15\Myr$ (panel {\it (c)} on the first row of Figure \ref{fig:emission2.4}). There is also a well-defined ring of bright emission defining the waist of the Bipolar Nebula, as seen in the face-on view at $\tOB\seq0.15\Myr$ (panel {\it (f)} on the second row of Figure \ref{fig:emission2.4}). Because the collision is slow, the \SCL\ breaks up relatively slowly, and is still quite coherent when the expanding Ionisation Front that defines the waist encounters it. Consequently the waist is approximately circular. Again because the collision is quite slow, when the \HIIR\ breaks out of the \SCL\ along the $x$ axis, there is still incoming, relatively dense cloud gas to be ionised, and therefore the lobes are bright.

Following the \hv\ collision the evolution of the \HIIR\ is faster, and the morphology is rather less regular. Small bipolar lobes have appeared by $\tOB\seq0.05\Myr$ (panel {\it (g)} on the third row of Figure \ref{fig:emission2.4}), but they subsequently disappear, and there is no obvious bipolarity in the maps at $\tOB\seq0.10\Myr$ and $0.15\Myr$ (panels {\it (h)} and {\it (i)} on third row of Figure \ref{fig:emission2.4}). The ring defining the waist is also starting to break up at $\tOB\seq0.10\Myr$ (panel {\it (k)} on the fourth row of Figure \ref{fig:emission2.4}) and is very fragmented by $\tOB\seq0.15\Myr$ (panel {\it (l)} on the fourth row of Figure \ref{fig:emission2.4}). Because the collision is fast, by the time the \HIIR\ breaks out of the \SCL, there is very little dense cloud gas left outside the \SCL\ to be ionised, so the lobes are very weak. And because the \SCL\ forms and breaks up (into a \SWS) very quickly, the expanding Ionisation Front at the waist meets very patchy resistance from the \SCL, which is why the waist is so fragmented.

%%%%%
\subsubsection{The evolution of the waist}
%%%%%

When viewed from directions close to the collision axis, the distinctive feature of an intrinsically bipolar \HIIR\ is not the two extended lobes of diffuse H{\sc ii} (because these are lined up behind, and in front of, each other), but the bright rim at the waist, where the ionising radiation meets and erodes the dense gas of the \SCL. The persistence of the waist determines the longevity of the bipolar morphology. Once the dense \SCL\ is fully eroded, the waist disappears and the \HIIR\ transitions to a more isotropic configuration, with little evidence of bipolarity.

Using maps of Emission Measure from face-on (viewing angle, $\theta\seq0^\circ$), like those on the second and fourth rows of Figure \ref{fig:emission2.4}, we can trace the evolution of the waist, as characterised by its radius, $\RW$. To locate the waist we define $\Ntot\seq250$ equally spaced radial rays from the origin of coordinates on the observer's sky, $[X,Y]$; each ray $n$ is distinguished by the angle, $\phin\seq2\pi n/\Ntot$, that it makes with the $X$ axis. We then scan outwards along each ray to find the maximum Emission Measure, $\etan$, the mean Emission Measure, $\mun$, and the standard deviation of the Emission Measure, $\sign$, on that ray. If the maximum is `significant', i.e. $\etan>\mun+10\,\sign$, we mark the ray as `significant' and note the distance, $\RWn$, between this maximum and the origin of coordinates. $\RW$ is then set to the mean of the `significant' $\RWn$ values

Fig.\ref{fig:waist} shows the evolution of the waist radius as a function of time. Red and blue curves correspond to collision velocities $2u_0=2.4$ and $4.0$ km s$^{-1}$, respectively. Solid, dash-dotted, and dashed lines represent sonic, supersonic, and hypersonic initial turbulence. 

\Hv\ collisions produce waists that expand rapidly, and the \SCL\ is fully ionised within $\sim0.20$ Myr. In contrast, \lv\ collisions produce waists that expand more slowly, and the \SCL\ survives until $\sim0.35$ Myr. Thus survival of the \SCL\ is primarily controlled by the collision velocity. Turbulence has a smaller effect on the survival of the \SCL. With lower turbulence, stars form more quickly, but the \SCL\ is more homogeneous and presents stiffer resistance to the advance of the \IF\ and the expansion of the waist.

%%%%%
\subsubsection{Viewing angle}\label{SEC:ViewingAngle}
%%%%%

Evidently the observed morphology of a Bipolar \HIIR\ depends critically on the viewing angle. The top row of Figure \ref{fig:viewing} shows Emission-Measure maps at $\tOB\seq0.15\Myr$ for a Bipolar \HIIR\ produced by a \lv\ collision ($2\uO\seq2.4\kms$) between clouds with sonic turbulence ($\Mbar\seq1$). From left to right the different maps show a sequence of viewing angles, starting edge-on to the \SCL\ ($\theta\seq90^\circ$; panel {\it (a)}) and progressing to directions closer to face-on ($\theta\seq75^\circ$, $60^\circ$ and $30^\circ$; respectively panels {\it (b)}, {\it (c)} and {\it (d)}); the exactly face-on view ($\theta\seq0^\circ$) is the righthand map on the second row of Figure \ref{fig:emission2.4} (panel {\it (f)}).

Even if the bipolarity of an \HIIR\ is well defined in 3D, as in the setup illustrated on the top row of Figure \ref{fig:viewing}, it is only clear that the \HIIR\ is bipolar if it is viewed at a large angle, $\theta$, to the collision axis. As $\theta$ decreases, the two lobes become ever more aligned with one another along the line of sight. The division between the two lobes (where the \SCL\ intervenes) becomes steadily more indistinct, and the overall \HIIR\ looks increasingly spherical.

At the same time the Bright Rim marking the waist opens out into an ellipse, which becomes increasingly circular as $\theta$ decreases. From a purely morphological perspective, at small $\theta$ the \HIIR\ looks like a classical, approximately spherical Monopolar \HIIR, limb-brightened in directions where it has swept up a dense shell, but with the illusion that some ionising radiation has escaped beyond this shell (see Section \ref{SEC:DiffEm}).

Thus the probability of observing this rather regular Bipolar \HIIR\ from a viewing angle greater than some critical value, $\thCRIT$, and therefore perceiving it as bipolar, is $P(\thCRIT)\seq\cos(\thCRIT)$, which for $\theta\!\sim\!60^\circ$ ({\it vice} $75^\circ$) is $\sim\!50\%$ ({\it vice} $\sim\!25\%$). Combined with the fact that the bipolar phase is quite short-lived, the chance of an \HIIR\ being seen as bipolar is low, even if it has, at some stage, gone through an unambiguously bipolar phase.

%%%%%
\subsubsection{Extended Diffuse Emission}\label{SEC:DiffEm}
%%%%%

The rather regular Bipolar \HIIR\ produced by a \lv\ collision involving clouds with sonic turbulence ($2\uO\seq2.4\kms$, $\Mbar\seq1$) has considerable extended diffuse Emission Measure on scales large than the Bright Rim defining its waist. This extended diffuse emission is visible from all angles, as shown on the upper row of Figure \ref{fig:viewing} (panels {\it (a)} to {\it (d)}). Most of this extended diffuse Emission Measure derives from the bipolar lobes, which spread out further from the collision axis than the waist, and are visible even at viewing angles close to the collision axis (e.g. $\theta\seq30^\circ$, panel {\it (d)} on Figure \ref{fig:viewing}; ~and $\theta\seq0^\circ$, panel {\it (f)} on Figure \ref{fig:emission2.4}). If the turbulent Mach Number ($\Mbar$) is increased, the Emission-Measure maps are less regular, but the outer reaches of the \HIIR\ are still defined by this extended diffuse Emission Measure from the bipolar lobes (for example, see Figure \ref{fig:viewingturb}). Whilst this diffuse emission is not an unambiguous signature of bimodality, it is circumstantial evidence.

If the collision velocity ($2\uO$) is increased, the waist is less well defined, and the diffuse emission beyond the fragmented bright rims on Figures \ref{fig:viewing4_0} and \ref{fig:viewing4_0turb} could be attributed to ionising radiation escaping through holes in a classical spherical bubble.

%%%%%
\subsubsection{Kinematics}\label{SEC:Kin}
%%%%%

Radio recombination-line emission provides information on the kinematics of ionised gas. Here we concentrate on the Emission-Measure weighted mean radial velocity, $\muv$ and the Emission-Measure weighted radial velocity dispersion, $\sigv$, but in principle more sophisticated signatures can be derived \citep[e.g.][]{whitworth2018bipolar}. The kinematic signatures of bimodality depend strongly on the viewing angle, $\theta$, and tend to be clearest when the level of turbulence in the pre-collision clouds is low.

The lower row of Figure \ref{fig:viewing} shows maps of $\muv$, for the \lv\ setup with sonic turbulence ($2\uO\seq2.4\kms,\;\Mbar\seq1$), i.e. a relatively regular Bipolar \HIIR. At large viewing angle (looking approximately edge-on to the \SCL), there is little systematic variation in $\muv$, as expected -- and as shown on panel {\it (e)} on the lower row of Figure \ref{fig:viewing}. As the viewing angle is reduced (progressing from left to right through the sequence of maps on the lower row of Figure \ref{fig:viewing}), a clear asymmetry emerges, because one lobe is approaching the observer (here, the mainly blue lefthand lobe), and the other lobe is receding from the observer (here, the mainly red righthand lobe). This asymmetry persists over a range of viewing angles. However, at small viewing angles ($\theta\!\lesssim\!30^\circ$), the asymmetry declines because the approaching blue lobe becomes increasingly well aligned with the receding red lobe, and the opposing radial velocities tend to cancel out giving low $\muv$.

A similar pattern is observed if the collision velocity is increased to $2\uO\seq4.0\kms$, as shown on Figures \ref{fig:viewing4_0} and \ref{fig:viewing4_0turb}. If the turbulece in the pre-collision clouds is increased to hypersonic ($\Mbar\seq6$), the velocity field is more disordered, but there is still a clear signature for intermediate viewing angles.

This asymmetry is more convincing on the plots of $\fPV(\tOB)$ on Figure \ref{fig:pv}. ~$\fPV$, is the fraction of pixels for which the Emission Measure weighted radial velocity has the same sign as its $X$ coordinate; ~$\tOB$ is the time since the formation of the first OB star. Results are shown for three different viewing angles, $\theta\seq75^\circ$, $45^\circ$ and $15^\circ$. The two lefthand panels {\it ((a)} and {\it (b)}) show results for \lv\ collisions, and the two righthand panels {\it ((c)} and {\it (d)}) show results for \hv\ collisions. Within each velocity pair, the lefthand map shows results obtained with sonic turbulence (panels {\it (a)} and {\it (c)}), and the righthand map shows results obtained with hypersonic turbulence (panels {\it (b)} and {\it (d)}).

For most setups there is a clear and persistent asymmetry parameter that reflects which lobe is directed away from the observer, and which is directed towards the observer. The exceptions are the collisions with hypersonic turbulence viewed at large $\theta$; in these setups the residual bulk velocities from the turbulence are comparable with the projected systematic velocities from the large-scale colliding flows, and the $\fPV$ signature is blurred.

Thus, in most cases the asymmetry parameter, $\fPV$, provides a useful signature supporting -- if not confirming definitively -- the identity of Bipolar \HIIR s at intermediate viewing angles, i.e. at viewing angles that are sufficiently large that the Bright Rim defining the waist may not present a clear ring, or are sufficiently small that the bipolar lobes are not clearly separated.

%%%%%
\begin{figure}
\centering
\includegraphics[width=1\linewidth]{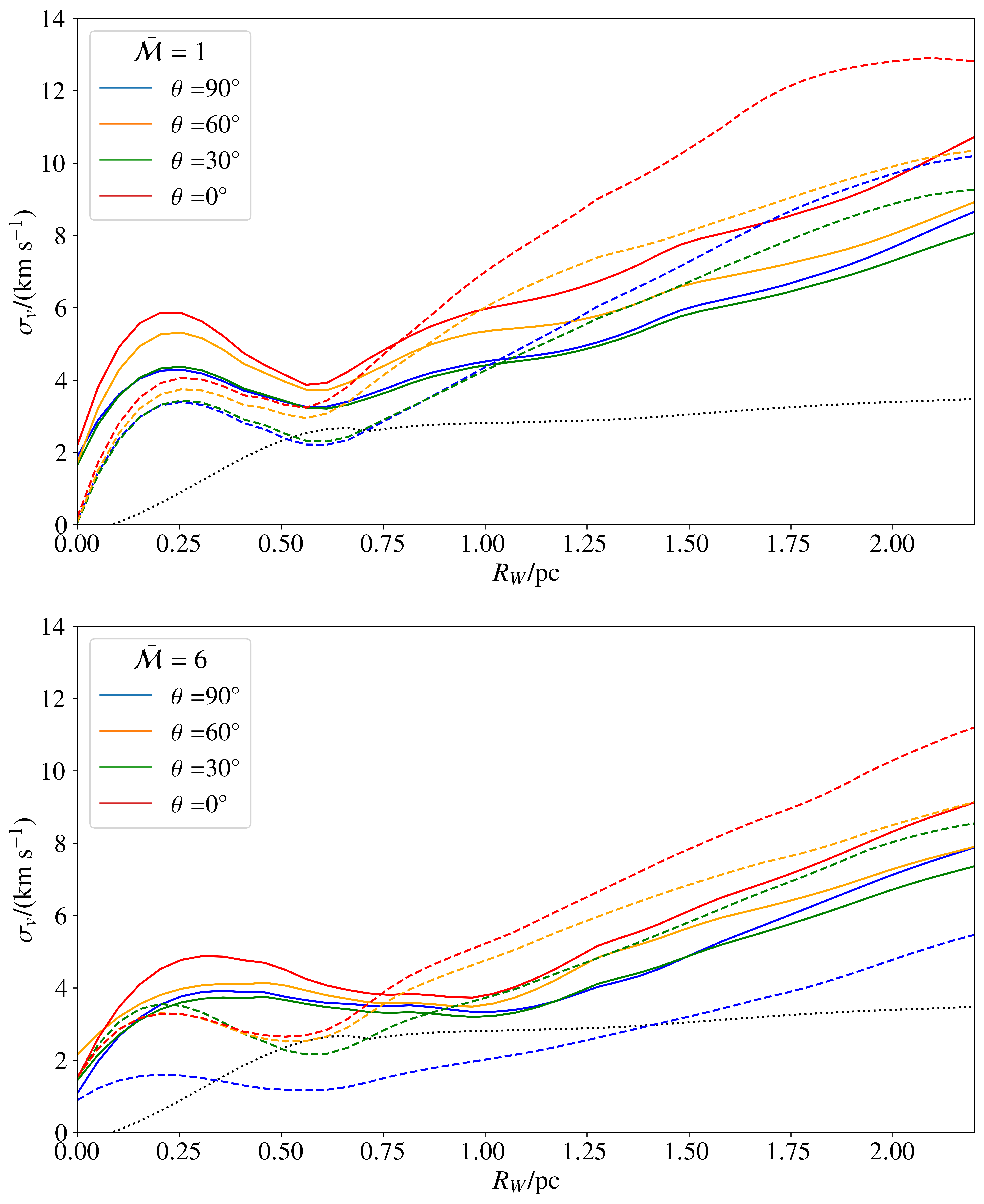}
\vspace{-0.3cm}
\caption{Line-of-sight velocity dispersion $\sigv$ as a function of the waist radius $\RW$, used here as a proxy for time. Solid curves present \lv\ collisions. Dashed curves present \hv\ collisions. Line colours represent viewing angles $\theta\seq0^\circ$ (blue), $30^\circ$ (orange) and $60^\circ$ (green). The dotted black curve presents the results from the Fiducial Setup. ~{\it Top Panel:} sonic turbulence, $\Mbar\seq1$. {\it Bottom Panel:} hypersonic turbulence, $\Mbar\seq6$. Each curve is the mean of two realisations}
\label{fig:mom}
\vspace{-0.45cm}
\end{figure}
%%%%%

Figure \ref{fig:mom} shows the radial velocity dispersion, $\sigv$, as a function of $\RW$ (used here as a proxy for time). As expected, $\sigv$ increases more-or-less monotonically with $\RW$. It reaches values $\sim\!5\kms$ to $\sim\!12\kms$ by $\RW\!\sim\!2\pc$, and is still continuing to increase. The rate of increase is higher for smaller viewing angles, but even at $\theta\seq60^\circ$ the values are high. There is little dependence on the level of turbulence. The Fiducial Setup delivers much smaller values of $\sigv$.

%%%%%
\begin{figure*}
\centering
%\vspace{0.3cm}
\begin{overpic}[width=1\linewidth]{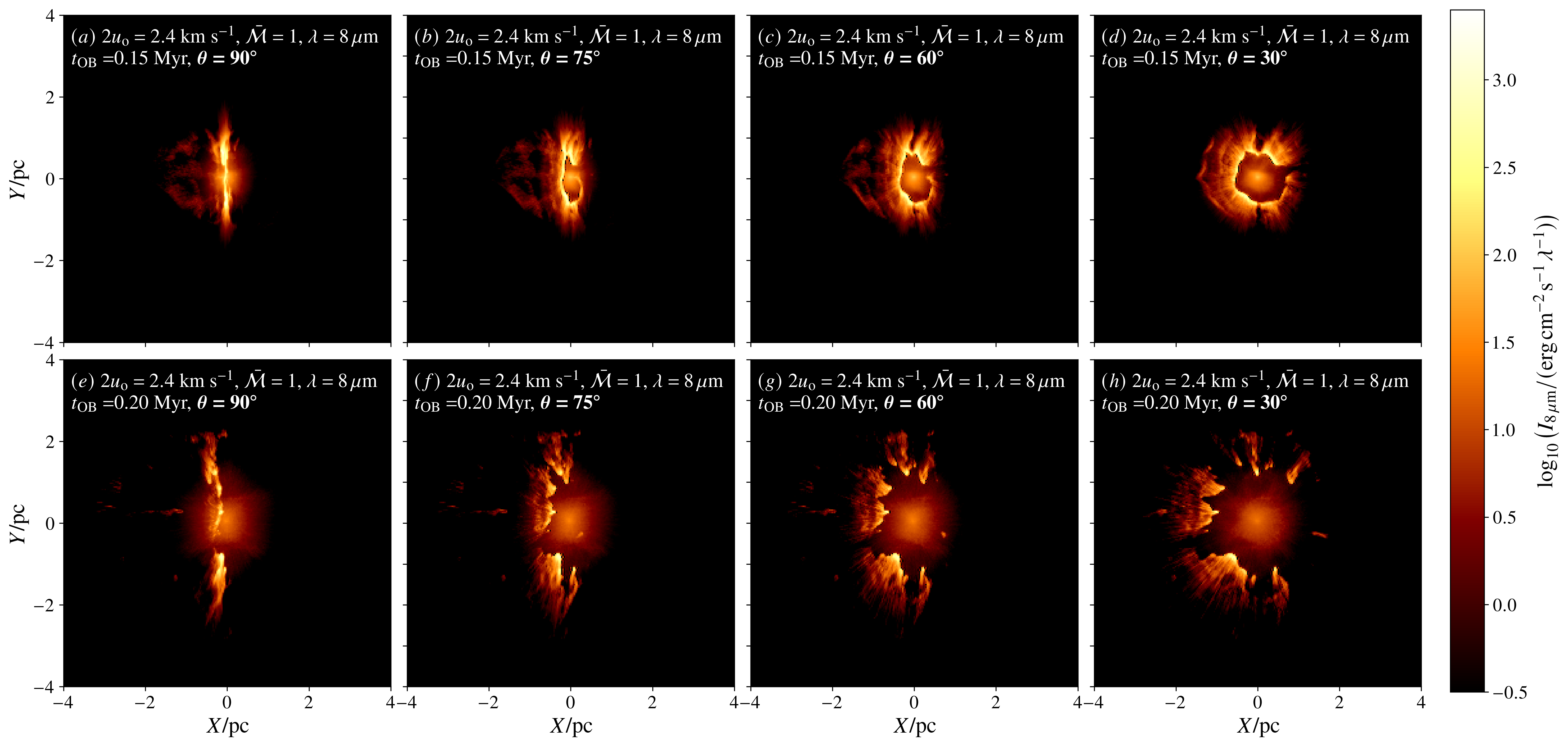}
%\put(0,250){\textbf{\eightmum}}
\end{overpic}\vspace{0.4cm}
\begin{overpic}[width=1\linewidth]{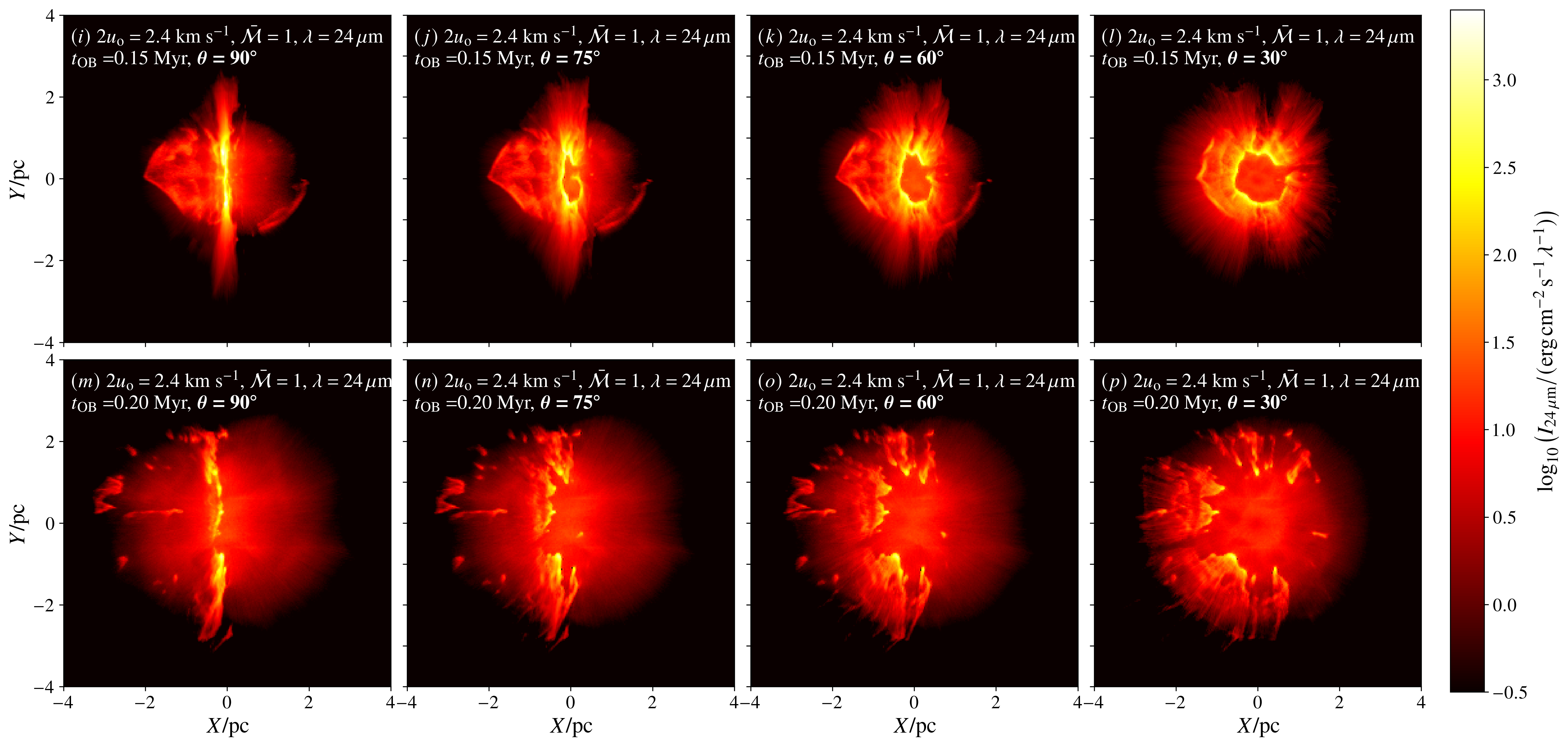}
%\put(0,250){\textbf{\twofourmum}}
\end{overpic}
\vspace{-0.6cm}
\caption{False-colour maps of dust emission at \eightmum\ (top two rows, panels {\it a} to {\it h}) and $24\,\mu\mathrm{m}$ (bottom two rows, panels {\it i} to {\it p}) for a \lv\ collision with sonic turbulence (the same setup as shown on the top two rows of Figures \ref{fig:ColDen} and \ref{fig:emission2.4}, and both rows of Figure \ref{fig:viewing}). The maps have been computed using \textsc{radmc‑3d} \citep{dullemond2012} and \textsc{dinamo} \citep{2019Felix}. From left to right the four columns represent decreasing viewing angle: $\theta\seq90^\circ$, $\,70^{\circ}$, $\,60^\circ$ and $\,30^\circ$. The first and third rows (panels {\it a} to {\it d}, and {\it i} to {\it l}) correspond to time $\tOB\seq0.15\Myr$, and the second and fourth rows (panels {\it e} to {\it h}, and {\it m} to {\it p}) correspond to $\tOB\seq0.20\Myr$. The values of $2\uO,\,\Mbar,\,\tOB,\,\lambda\,{\rm and}\,\theta$ are given at the top of each panel. See the text for details of the mixture of dust species invoked.}
\label{fig:dustmaps}    
\end{figure*}
%%%%%

%%%%%
\subsection{Mid-infrared observations of dust}\label{SEC:MidIRDust}
%%%%%

The structure of an \HIIR\ can be traced using the strong mid-infrared emission at \eightmum\ and \twofourmum\ from Polycyclic Aromatic Hydrocarbons (PAHs) and other dust species, especially the strongly heated dust in the \PDR\ just behind the \IF\ \citep[e.g.][]{deharveng2015bipolar}.

To enable a comparison with such observations, we post-process our experiments using \textsc{radmc-3d} \citep{dullemond2012}, on the assumption that the dust is a mixture of silicate, carbonaceous, and PAH dust grains, with their densities depending on the gas density of each grid cell $\rho$ as
\begin{eqnarray}
\left.\begin{array}{rcl}
\rho_{\mbox{\tiny SILICATE}} &=& 0.0070\;\rho\;{\rm e}^{-\fion/1.5}\,,\\
\rho_{\mbox{\tiny CARBONACEOUS}} &=& 0.0021\;\rho\;{\rm e}^{-\fion/0.6}\,,\\
\rho_{\mbox{\tiny PAH}} &=& 0.0009\;\rho\;{\rm e}^{-\fion/0.3}\,.\\
\end{array}\right\}
\end{eqnarray}
Here $\fion$ is the fraction of ionised gas in a given cell, and the exponential factors reflect the survivability of different dust species in ionised environments  \citep{2016salgado,2011jones,2015Akimkin,2019Topchieva}. In fully neutral gas this prescription gives the standard dust to gas ratio of $0.01$.

The radiation field from \textsc{radmc-3d} is then sampled across UV to near-infrared wavelengths, and the resulting emission is computed using the \textsc{dinamo} framework \citep{2019Felix}, which corrects for stochastic heating of small grains. The resulting \eightmum\ maps are shown on the top two rows of Figure \ref{fig:dustmaps} (panels {\it (a)} to {\it (h)}); the corresponding \twofourmum\ maps are shown on the bottom two rows of Figure \ref{fig:dustmaps} (panels {\it (i)} to {\it (p)}). Rows 1 and 3 (panels {\it (a)} to {\it (d)}, and {\it (i)} to {\it (l)}) correspond to $\tOB\seq0.15\Myr$; rows 2 and 4 (panels {\it (e)} to {\it (h)}, and {\it (m)} to {\it (p)}) correspond to $\tOB\seq0.20\Myr$. From left to right, the four columns represent decreasing viewing angles $\theta\seq90^\circ\!$, $75^\circ\!$, $60^\circ\!$ and $30^\circ\!$.

%%%%%
\subsubsection{$\;8\,\upmu{\rm m}$ emission maps}\label{SEC:8um}
%%%%%

PAH grains are small, and they are quickly destroyed by ionising radiation, so they only survive near the boundaries of, and outside, \HIIR s \citep{2023Egorov}. Due to their small heat capacity, the PAH grains near the boundary of an \HIIR\ are heated stochastically to high temperatures by individual photons in the intense UV radiation field, and consequently they dominate the \eightmum\ emission from these regions \citep{2007draine}. Thus, maps of \eightmum\ emission at $\tOB\seq0.15\Myr$ from small viewing angles (panels {\it (c)} and {\it (d)} on Figure \ref{fig:dustmaps}) trace the waist of the Bipolar \HIIR. Even when the waist has started to break up at $\tOB\seq0.20\Myr$ (panels {\it (g)} and {\it (h)} on Figure \ref{fig:dustmaps}), the location of the waist is still evident from small viewing angles.

In contrast, on maps at large viewing angles the \eightmum\ emission is concentrated in a narrow linear structure delineating the waist seen edge-on, and does not show bipolar lobes. This is  because the lobes are almost devoid of dust, and in particular of PAH dust. Thus, the morphological classification of \HIIR s based on \eightmum\ emission is inherently and systematically biased against recognising Bipolar \HIIR s.

%%%%%
\begin{figure}
\centering
\includegraphics[width=1\linewidth]{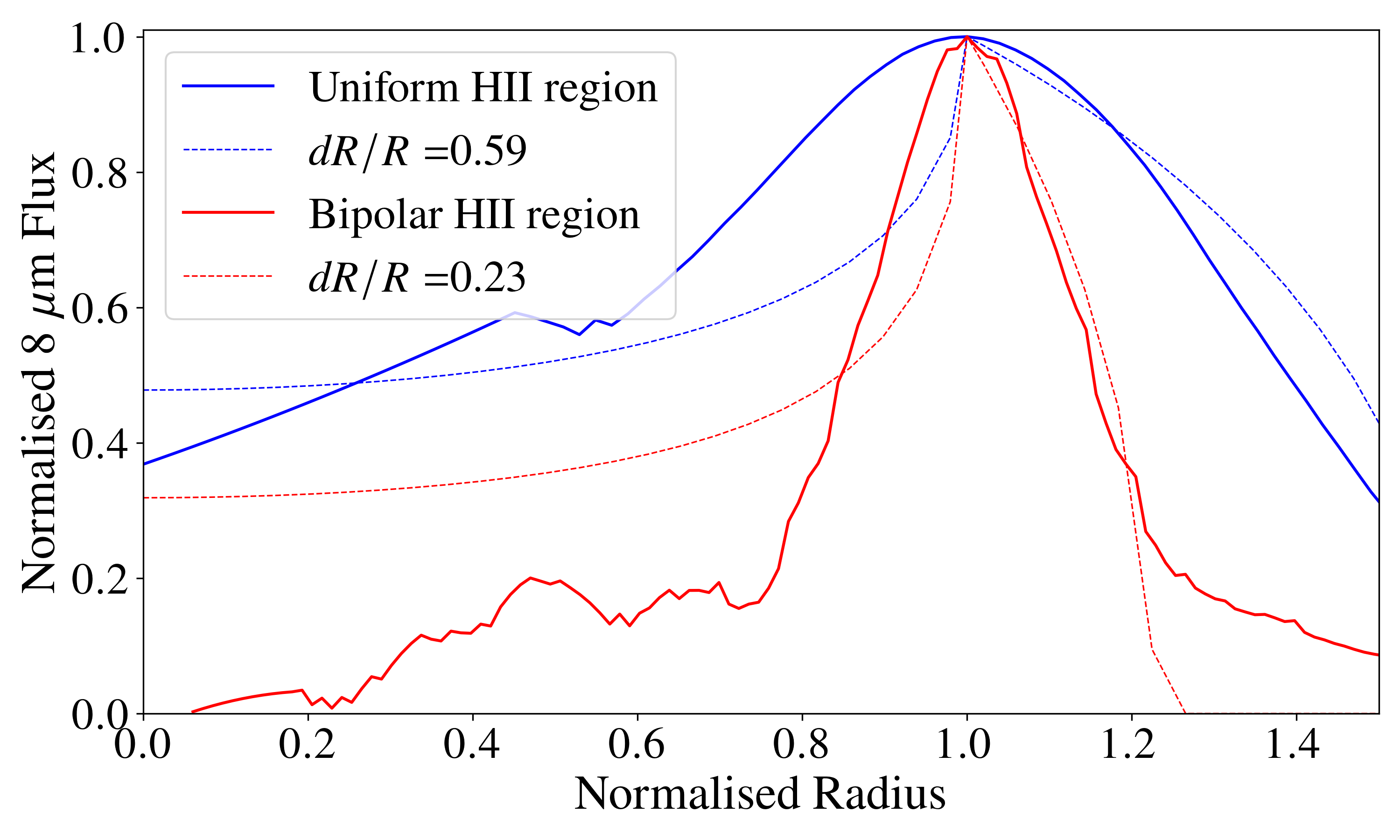}
\vspace{-0.3cm}
\caption{Normalised mean radial profiles of \eightmum\ emission, i.e. $I(b)/I(R)$ against $b/R$. The full red line shows the mean radial profile for the two realisations of the Bipolar Setup with $2\uO\seq2.4\kms$ (\lv) and $\Mbar\seq1$ (sonic turbulence), viewed at $\theta\seq60^\circ$. The full blue line is the mean radial profile for the spherically symmetric Fiducial Setup. The dashed red and blue lines are the corresponding best-fit analytic profiles obtained using Equation \ref{EQN:Radial}. See text for further details.}
\label{fig:radialmicron}
\end{figure}
%%%%%

The strength of \eightmum\ emission from the waist, combined with the weakness of \eightmum\ emission from the lobes, results in very low \eightmum\ emission inside the waist at small viewing angles. At first glance, this might be attributed to limb-brightening in a classical, approximately spherical bubble. If we model the \eightmum\ emission from a classical, approximately spherical \HIIR\ as arising solely from a shell between radius $R$ and radius $R\!+\!\Delta R$, and if we assume uniform, optically thin emission from this shell, the intensity as a function of impact parameter $b$ is
\begin{eqnarray}\label{EQN:Radial}
I(b)\!\!\!&\!\!\!=\!\!\!&\!\!\!\!\left\{\!\!\!\begin{array}{ll}
\IO\!\left\{\!\left(\left[\!1\!+\!\frac{\Delta R}{R}\!\right]^{\!2}\!-\!\left[\!\frac{b}{R}\!\right]^{\!2}\right)^{\!1/2}\!-\!\left(1\!-\!\left[\!\frac{b}{R}\!\right]^{\!2}\right)^{\!1/2}\!\right\}\!,&\!\!\!\!\!b\!\leq\!R,\\
\IO\!\left(\left[\!1\!+\!\frac{\Delta R}{R}\!\right]^{\!2}\!-\!\left[\!\frac{b}{R}\!\right]^{\!2}\right)^{\!1/2}\!,&\!\!\!\!\!R\!<\!b\!<\!R\!+\!\Delta R,\\
0,&\!\!\!\!\!b\!\geq\!R.\\
\end{array}\!\!\right.\hspace{0.25cm}
\end{eqnarray}
The intensity therefore has a maximum at $b\seq R$ (looking along the inside surface of the shell), and a minimum at $b\seq0$ (looking through the middle of the shell). The profile can be characterised by two fundamental parameters: the width of the maximum at $b\seq R$, which is $\sim\!\Delta R/2$, and the depth of central minimum,
\begin{eqnarray}
{\cal D}\;\;\,\equiv\,\;\;\frac{I(0)}{I(R)}&=&\left\{\frac{2R}{\Delta R}\,+\,1\right\}^{-1/2}\,.
\end{eqnarray}
The thinner the shell is, the narrower is the maximum, {\it and} the deeper is the central minimum relative to the maximum (i.e. small ${\cal D}$). 

The red line on Figure \ref{fig:radialmicron} shows mean intensity profile from the two realisations of a CCC at $2\uO\seq2.4\kms$, involving clouds with sonic turbulence ($\Mbar\seq1$) (hereafter the approximately cylindrical Bipolar Setup), at $\tOB\seq0.15\Myr$. The blue line shows the mean intensity profile from the Fiducial Setup, when the shell radius has reached the same radius as the Bipolar Setup. The impact parameter, $b$, is measured from the centre of coordinates on the observer's sky ($[X,Y]\seq[0,0]$), and the intensity is averaged azimuthally relative to this point. $R$ is the impact parameter at which the mean intensity is maximum.

The dashed lines on Figure \ref{fig:radialmicron} are the corresponding best fits, obtained using Equation \ref{EQN:Radial} with $\Delta R/R\seq0.59$ (blue dashed line; best fit to the approximately spherical Fiducial Setup) and $\Delta R/R\seq0.23$ (red dashed line; best fit to the approximately cylindrical Bipolar Setup). The deep central minimum in the mean radial intensity profile delivered by the approximately cylindrical Bipolar Setup cannot be obtained with Equation \ref{EQN:Radial}, if the width of the maximum defining the waist is also reproduced. Thus, at least in the context of the \HIIR s modelled here, the mean radial intensity profile can be used as a diagnostic of Bipolar \HIIR s seen at small viewing angles.

Whether this diagnostic is applicable to real observations depends on the level of background confusion. The same methodology has been applies to CO obervations of the molecular gas swept up at the boundaries of \HIIR s, to infer that they are often cylindrical \citep[e.g.][]{Beaumont2010}.

%%%%%
\subsubsection{\twofourmum\ emission maps}
%%%%%

At \twofourmum\ the emission includes contributions from larger grains that are significantly more resilient to destruction and can survive in ionised gas. Consequently, the bipolar lobes can be seen clearly on \twofourmum\ maps at large viewing angles, and there is significant \twofourmum\ emission inside the waist at small viewing angles.

Taken together, the \eightmum\ and \twofourmum\ maps provide complementary diagnostics for intrinsically bipolar \HIIR s. The \eightmum\ emission traces the waist most clearly at small viewing angles, and the \twofourmum\ emission traces the bipolar lobes at large viewing angles.

%%%%%
\section{Discussion}\label{sec:Disc}
%%%%%

In the Appendix to \citet{balfour2017} we have shown that, statistically, all massive star formation {\it could} be triggered by CCCs. In that analysis, the dominant contribution to massive star formation is not from CCCs {\it between} large-scale Giant Molecular Cloud Complexes, but rather from CCCs between the intermediate-scale clouds of a fractal hierarchy of clouds {\it within} large-scale Molecular Cloud Complexes. This is because collisions involving intermediate-scale clouds, colliding at modest relative velocities (as modelled here), are more frequent than faster collisions between Giant Molecular Cloud Complexes. Provided that the collision velocity is not too high, $2\uO\!<\!\duC\!\simeq\!3.2\kms$, these collisions between intermediate-scale clouds naturally produce Hub-Filament Systems, with the filaments feeding material into a central Hub, where a monolithic cluster then forms.

%%%%%
\subsection{Magnetic fields}
%%%%%

In \citet{georgatos}, we have shown that the inclusion of a magnetic field, aligned with the collision velocity, does not significantly affect this mechanism for forming Hub-Filament Systems. The magnetic field simply increases the critical collision velocity, $\duC$, below which \HFS s form, and thereby increases the frequency with which CCCs form monolithic star clusters. It does not alter the underlying pattern of star formation. The \SCL\ still fragments into filaments, which are then dragged into radial orientations. The distribution of filament widths, and the relationship between filament axes and the associated magnetic field lines, are both in good agreement with observation.\footnote{ We note that in the simulations of isolated clouds presented by \citet{SuinPetal2025AA698A119}, a \HFS\ is produced with a relatively weak magnetic field, and a \SWS\ is produced with a stronger magnetic field. In contrast, our experiments produce \HFS s more readily with a stronger magnetic field. This apparent conundrum is mainly attributable to the different setups (an isolated single cloud, as opposed to a CCC), and the fact that the mass-to-flux ratio of the \SCL\ produced by the colliding clouds is larger. As a consequence the \SCL\ contracts laterally, and this is what pulls the filaments into radial orientations, thereby feeding matter into the central Hub.}

Since it appears that the inclusion of a magnetic field does not hugely affect the pattern of star formation triggered by a CCC, but simply extends the parameter space required to produce \HFS s, and since we are minded to keep the parameter space small, we have not included a magnetic field in the numerical experiments reported here. We reiterate that they are numerical experiments, rather than simulations, because the aim is to isolate and understand particular causes and effects, rather than to reproduce all the features of a real, and therefore chaotic, \HFS s. 

In the same spirit, we have simplified the initial conditions by (a) invoking uniform-density clouds with the same mass and radius; (b) giving the internal turbulent velocity field a thermal mix of solenoidal and compressive modes; and (c) setting the two clouds to collide head-on. Sink particles are introduced at a relatively low-density, $\rhosink\seq10^{-19}\gccc$ (equivalent to $2\times 10^4\,{\rm H}_2\ccc$), and each sink represents a group of stars with a random mass distribution drawn from a prescribed IMF, so the formation of individual stars is not resolved. Further work will be required to determine how the outcomes of CCCs might be altered if these simplifications were relaxed.

%%%%%
\subsection{Cloud internal substructure}
%%%%%

In this paper we have only explored the effects of different levels of internal substructure in the pre-collision clouds, and the effects of different collision velocities. Moreover, the consequences of internal substructure might have been better represented if there had been, from the outset, density substructure consistent with the turbulent velocity field -- rather than the density substructure only developing as a consequence of a prescribed turbulent velocity field. The physics that {\it is} followed is the self-gravitating hydrodynamics of the formation and fragmentation of a \SCL; the formation of a \HFS\ and the condensation of sinks; the concentration of sinks in the central Hub and feedback from the sinks; the chemistry and thermal physics of both the neutral and ionised gas; and the associated radiation transport.

%%%%%
\subsection{Bipolar \HIIR s and Cloud-Cloud Collisions}\label{SEC:BPHIIR.CCC}
%%%%%

The main results from the numerical experiments are (i) that relatively slow collisions between intermediate-mass clouds can trigger the formation of a \HFS, and the formation of a star cluster that includes massive stars ($M_\star\!\gtrsim\!10\Msun$); (ii) that ionising radiation from the massive stars can, {\it under the right circumstances}, produce a bipolar \HIIR; (iii) that, seen from a sufficiently large angle to the collision axis, the two ionised lobes of the bipolar \HIIR\ are clearly visible; (iv) that, seen from a sufficiently small angle to the collision axis, the bright rim at the waist of the bipolar \HIIR\ could be mistaken for a classical, approximately spherical, limb-brightened H{\sc ii} bubble; (v) that there are geometric and kinematic signatures that might be used to  distinguish between the waist of a bipolar \HIIR\ viewed along the collision axis, and a classical H{\sc ii} bubble. 

Since the role of CCCs in triggering star formation is hard to estimate observationally, it is interesting to consider whether Bipolar \HIIR s might be circumstantial evidence for the importance of CCCs. There are two fundamental questions here. Firstly, are CCCs the most likely explanation for Bipolar \HIIR s? Secondly, are sufficient Bipolar \HIIR s observed to support the idea that CCCs are a dominant star formation trigger? We suggest that the answer to both questions is affirmative.

Firstly, it is difficult to explain the formation of a Bipolar \HIIR\ unless the setup involves a favoured axis defining the direction of the displacement between the two lobes. Thus an alternative setup for the formation of a Bipolar \HIIR s might be an isolated ellipsoidal cloud which collapses to a pancake; stars, including massive ones, then form at the centre and excite an \HIIR; and the \HIIR\ blows out along the direction of least resistance, i.e. orthogonal to the plane of the pancake. The ellipsoidal cloud might have been built up by smaller clouds colliding and merging, but the non-spherical geometry that results in collapse to a pancake would be attributable to the stochastic nature of the inflows assembling a non-spherical cloud. However, the velocity of the material accreting onto the pancake would now be attributable solely to the cloud's self-gravity, and would therefore be insufficient to form a \HFS; this in turn would make it much harder to form the massive stars required to excite an \HIIR.

Secondly, not all CCCs will produce a Bipolar \HIIR. On the basis of the experiments reported here, the formation of a well-defined Bipolar \HIIR\ requires a relatively slow collision between two relatively quiescent clouds. If the collision is too fast, or the level of turbulence in the clouds is too high, the \SCL\ is broad and disordered, and the waist of the Bipolar \HIIR\ is irregular, with prominences like {\it The Pillars of Creation} pointing towards the ionising stars (see Figure \ref{fig:dustmaps}, panels {\it (h)} and {\it (p)}), and less well collimated lobes (see Figure \ref{fig:emission2.4}, panel {\it (i)}). In addition, the two lobes will only be comparable in scale and intensity if the massive stars remain close to the $x\seq0$ plane, and the gas into which the two lobes expand, on either side of the \SCL, has similar density; otherwise there may only be one significant lobe.

Thirdly, not all Bipolar \HIIR s will necessarily be identified as such. Even if the bipolar symmetry of the \HIIR\ is well defined in 3D, this will only be obvious if the viewing angle (relative to the collision axis) is large. If it is not, subtler and more ambiguous morphological and kinematic signatures need to be evaluated, as discussed in Sections \ref{SEC:RadioHII} and \ref{SEC:MidIRDust}, for example: extended, but weak, freefree lobe emission from outside the bright rim defining the waist (see Section \ref{SEC:DiffEm} and Figure \ref{fig:emission2.4}, panel {\it (f)}); systematic velocity gradients and large velocity dispersion in radio recombination line emission (see Section \ref{SEC:Kin}, Figure \ref{fig:viewing}, panels {\it (f)} and {\it (g)}, Figure \ref{fig:pv} and Figure \ref{fig:mom}); a lack of \eightmum\ emission on lines of sight through the middle of the waist (see Section \ref{SEC:8um}, Figure \ref{fig:dustmaps}, panels {\it (d)}, {\it (l)}, Figure \ref{fig:radialmicron}, and Section \ref{SEC:ABW} below). 

Fourthly, the bipolar phase is transient. It does not start until the \HIIR\ breaks out of the \SCL, and it ends when the waist disintegrates and the \SCL\ is dispersed, or the lobes become too diffuse to detect. In the experiments reported here, the duration of the bipolar phase is typically only $\sim\!0.05\Myr$

%%%%
\subsection{Comparison with Bipolar \HIIR\ observations}
%%%%
Several observed Bipolar \HIIR s exhibit characteristics that are qualitatively similar to those produced in our numerical experiments. \cite{deharveng2015bipolar} performed a morphological study of G319.88+00.79 and G010.32$-$00.15 using a combination of near-IR and mid-IR observations. They identified peak column densities of $\sim10^{23}$ cm$^{-2}$ associated with the dense waist separating the two ionised lobes, with typical waist sizes of $\sim1$ pc. Although their study primarily focused on morphology and therefore provided limited kinematic information, the bipolar structure of these regions is clearly identifiable due to the favourable viewing angle between the dense waist and the observer.
Radio recombination line observations by \cite{veena2017} of G351.69$-$1.15 and G351.63$-$1.25 revealed velocity dispersions $\sigma_v$ of $\sim15$ km s$^{-1}$ and relative velocities of $\sim10$ km s$^{-1}$ between the two ionised lobes. Similarly, \cite{dalgleish2018} measured a velocity separation of $\sim15$ km s$^{-1}$ and $\sigma_v\sim8$ km s$^{-1}$ in G316.81$-$0.06.

Our numerical experiments display several similarities to these observational studies. At a waist size of $\sim 1$ pc, the  $24,\upmu$m emission exhibits bright emission from ionised lobes separated by a dense, coherent high emission waist, making the bipolar morphology readily identifiable at large viewing angles. The waist column densities are of the order $\Sigma\sim1$ g cm$^{-2}$, comparable to \cite{deharveng2015bipolar} observations. The simulated ionised gas kinematics are also broadly consistent with the available RRL observations, with relative velocities between the two lobes reaching $\sim20$ km s$^{-1}$. The velocity dispersion depends strongly on viewing angle, ranging from $\sigma_v\sim7$ km s$^{-1}$ at high viewing angles to $\sigma_v\sim14$ km s$^{-1}$ when viewed closer to the symmetry axis.

Given these similarities, it would be tempting to make direct comparisons between our results and observational studies. However, this would require extensive study of the parameter space, including, but not limited to, a larger cloud mass spectrum, inclusion of non head-on collisions, magnetic fields, and unequal mass clouds. 
Thus, we stress that the results presented here represent idealised numerical experiments with emphasis on isolating cause and effect by starting from a small parameter space. Agreement with observations should be interpreted as demonstrating that Bipolar \HIIR s produced by \CCC\ can potentially reproduce several of the observed characteristics of Bipolar \HIIR s, but should not be directly compared with observational studies. The expansion of the parameter space to enable direct comparison with observations is left for future work.

%%%%%
\subsection{Bubbles or rings?}\label{SEC:ABW}
%%%%%

Using \textit{Spitzer} \eightmum\ observations, \cite{Churchwell2006,Churchwell2007} have identified a large population of ring-like infrared bubbles throughout the Galactic plane. Subsequent surveys have shown that such structures are widespread in the ISM, and are commonly associated with embedded OB-type stars and expanding H{\sc ii} regions \citep{Simpson2012,Anderson2014,Deharveng2010,kumar2020,2025cygnus}. However, the intrinsic geometry of these bubbles remains debated.

Several studies have argued that some \textit{Spitzer} bubbles arise from \HIIR s expanding within flattened molecular clouds, producing projected ring-like morphologies rather than fully spherical shells \citep{Beaumont2010, Deharveng2010, Hou2014, deharveng2015bipolar, whitworth2018bipolar}. In particular, \citet{Beaumont2010} conclude, on the basis of CO observations, that many bubbles are more consistent with flattened ring-like structures, than with spherical shells, and \citet{Hou2014} find that the molecular gas surrounding \HIIR s is frequently distributed anisotropically. In contrast, \citet{Anderson2012}, based on the results of \citet{Bania2010} and \citet{Anderson2011}, and the relatively low population of Bipolar \HIIR s in the region studied, argue that most bubbles are consistent with expanding three-dimensional shells. More recent studies suggest that both interpretations may be valid, depending on environmental conditions and viewing angle \citep{Pabst2020, kumar2020}.

Our numerical experiments highlight that Bipolar \HIIR viewed at low-intermediate $\uptheta$ are capable of producing ring-like (bubble) morphologies in 8$\upmu$m maps. Their radial profiles could be misidentified as limb-brightened thin-shell spherical \HIIR. Consequently, radial emission profiles alone may not uniquely distinguish  between cylindrically or spherically symmetric intrinsic bubble geometries.  Bipolar \HIIR\ also produce broadened RRL profiles. This is consistent with the broad RRL FWHM values reported by \citet{Bania2010} for many ISM bubbles, while our spherically symmetric numerical experiment produced significantly smaller RRL profiles. These could indicate that, at least a subset of ISM bubbles may represent  Bipolar \HIIR.

%%%%%
\section{Summary}\label{SEC:Conc}
%%%%%

We have modelled collisions between uniform-density clouds, each with mass $500\Msun$ and radius $2\pc$. The clouds have internal turbulence, with a thermal mix of solenoidal and compressional modes, and mean Mach Number $\Mbar\seq1$, $3$ or $6$. They collide head-on with velocity $2\uO\seq2.4\kms$ or $4.0\kms$. For each combination of $\Mbar$ and $2\uO$ we run two realisations, with different seeds for the turbulent velocity field, giving a total of twelve CCC runs.

The collisions produce \SCL s, which fragment to produce \HFS s. Sinks condense out in the Hub, and grow by Competitive Accretion. The sinks represent small star clusters, with a mix of stellar masses drawn randomly from a prescribed mass function (rather than individual stars). Once the sink mass passes $120\Msun$, the mix includes massive stars ($M_\star\!\gtrsim\!10\Msun$). The radiation from these massive stars ionises the surrounding gas which quickly blows out of the \SCL\ along the direction of smallest column-density (i.e. the collision axis, defined by the pre-collision bulk velocities of the clouds), producing a Bipolar \HIIR.

In addition we have modelled a single stationary cloud, again with mass $500\Msun$ and radius $2\pc$, but now with a single sink particle at its centre, from the beginning. The sink particle has initial mass $120\Msun$, and so its stellar mix includes a massive star from the outset. Because the surrounding gas is distributed approximately isotropically, the resulting  \HIIR\ develops into a classical, approximately spherical H{\sc ii} bubble. This configuration is referred to as The Fiducial Setup, and serves as a reference for comparison with the Bipolar \HIIR s modelled in the other 12 runs. 

In 3D, the \SCL\ and Bipolar \HIIR\ are well defined provided that the colliding clouds are relatively quiescent and the collision is slow. If the clouds are more  turbulent and/or the collision is faster, the \SCL\ and the Bipolar \HIIR\ are more disordered: one lobe may be much more luminous and/or extensive than the other, and the bright rim defining the waist of the \HIIR\ may be very fractured.

In 2D (i.e. in projection, as seen by an external observer), the structure of the Bipolar \HIIR\ depends strongly on the viewing direction, relative to the collision axis. We distinguish two limiting cases, characterised by the angle, $\theta$, between the viewing direction and the collision axis.\\

\noindent If viewed from a small angle to the collision axis:
\vspace{-0.23cm}
\begin{itemize}
\item The \SCL\ is seen face-on, and the main observed feature is the bright rim marking the waist of the Bipolar \HIIR. This bright rim is particularly well defined in free-free emission and radio recombination line emission from the newly ionised gas being boiled off the rim.
\item The waist is also visible  in \eightmum\ emission from small PAH dust grains. These grains are transiently heated to very high temperatures in the intense UV radiation field near the \IF. The PAH emission from the neutral material behind the \IF\ is weaker because the UV radiation field there is weaker. And the PAH emission from the ionised material well in front of the \IF\ is weaker because PAHs do not survive long in ionised gas.
\item Because they are aligned along the line of sight, the bipolar lobes are not clearly separated on the sky. However, the expansion of the lobes towards, and away from, the observer gives a large radial velocity dispersion in recombination-line emission, much larger than the radial velocity dispersion from The Fiducial Setup.
\item Because the lobes not only expand along the collision axis, but also expand laterally away from the collision axis, there is low-intensity freefree and recombination-line emission that might be detectable outside the waist-ring. If detected, this extended emission would constitute further evidence for a Bipolar \HIIR\ -- as distinct from a classical, limb-brightened, radiation-bounded, bubble.
\end{itemize}

\noindent If viewed from a large angle to the collision axis:
\vspace{-0.23cm}
\begin{itemize}
\item The \SCL\ is seen edge on, and the main observational features are the high column-density of the edge-on \SCL, the strip of emission from the waist, and the two lobes (one on each side of the \SCL).
\item Whereas the column-density along the collision axis and through the initial setup is only $N\simeq4\times 10^{21}\,{\rm H}_2\cc$ (equivalently $\Sigma\seq0.02\gcc$), the column-density at large $\theta$ can reach values as high as $N\!\simeq\!10^{24}\,{\rm H}_2\cc$, due to compression and projection. The high column-density of the edge-on \SCL\ is visible in dust obscuration at short wavelengths, and in dust emission at long wavelengths. It is also visible in molecular-line emission.
\item The Bright-Rim defining the waist of the Bipolar \HIIR\ is seen in strong freefree-continuum and recombination-line emission at radio wavelengths, and in \eightmum\ dust-continuum emission. Because the waist is aligned with the \SCL\ (in the sense that it is a thin, approximately circular ring close to the $x\seq0$ plane), it is seen edge-on and appears as a linear strip of intense emission.
\item The lobes are visible in freefree and recombination-line emission at radio wavelengths. This emission may even be visible at shorter wavelengths if foreground dust obscuration is low.
\item A lobe might appear to be a classical H{\sc ii} Bubble, strongly {\it radiation-bounded} and with a bright rim in the directions towards the \SCL, but possibly {\it density-bounded} in other directions. This interpretation would be particularly tempting if the massive stars produced in the Hub ended up concentrated on one side of the \SCL, so that one lobe was significantly larger and brighter than the other. 
\end{itemize}

%%%%%
\section{Conclusions}
%%%%%

We conclude:\\
{\it (A)} that Cloud-Cloud Collisions are a natural and compelling way to produce Bipolar \HIIR s, and therefore, {\it de facto}, a natural and compelling way to produce massive stars; they do so by
\vspace{-0.26cm}
\begin{enumerate}
     \item creating a \SCL\ which fragments into  a \HFS;
     \item the \HFS\ then concentrates mass into the central Hub;
     \item and this creates the conditions needed for massive stars to form;
\end{enumerate}
\vspace{-0.21cm}  
\noindent {\it (B)} that
\vspace{-0.26cm}
\begin{enumerate}
     \item since only a limited range of collisions parameters (quiescent clouds, slow collisions) produce Bipolar \HIIR s,
     \item since Bipolar \HIIR s can easily be mistaken for classical limb-brightened H{\sc ii} Bubbles,
     \item and since the Bipolar phase is short-lived,
\end{enumerate}
\vspace{-0.21cm}
\noindent Cloud-Cloud Collisions may play a critical role in triggering the formation of most massive stars, even though clearly identified Bipolar \HIIR s are only a small fraction of all \HIIR s.

%%%%%
\section*{Acknowledgements}
%%%%%

TG gratefully acknowledges the receipt of an STFC PhD studentship.
RW acknowledges the support by the Czech Ministry of
Education, Youth and Sports, through the INTER-EXCELLENCE II
program, project LUC24023 (MSMT-14950/2024-5), and by the
project RVO:67985815.

%%%%%
\section*{Data Availability}
%%%%%
 
Data is available from TG, upon request.

% REFERENCES %
\bibliographystyle{mnras}
\bibliography{example} % if your bibtex file is called example.bib
%%%%%

% CLOSE %
\bsp	% typesetting comment
\label{lastpage}
\end{document}